\documentclass[showpacs,showkeys,eqsecnum,aps,prc,floatfix]{revtex4}

\usepackage[dvips]{graphicx,color}
\usepackage{bm}

\begin{document}

\title{Influence of the \boldmath$N^\ast(1440)$ and \boldmath$N^\star(1535)$ 
Resonances in Intermediate Energy \boldmath$pp$ and \boldmath$np$ Scattering}

\author{A. Pricking$^{(a,b)}$, Ch. Elster$^{(c)}$, A. G{\aa}rdestig$^{(d)}$, F.
Hinterberger$^{(a)}$ \\
and the EDDA Collaboration}
\affiliation{$^{(a)}$Helmholtz Institut f\"ur Strahlen- und Kernphysik,
University of Bonn, Nussallee 14-16,  D-53115 Bonn, Germany}
\affiliation{$^{(b)}$Physikalisches Institut der Universit\"at T\"ubingen, D-72076
T\"ubingen, Germany} 
\affiliation{
$^{(c)}$Institute of Nuclear and Particle Physics,  and
Department of Physics and Astronomy,  Ohio University, Athens, OH 45701, USA}
\affiliation{$^{(d)}$Department of Physics and Astronomy, University of South Carolina,
Columbia, SC 29208, USA}

\vspace{10mm}

\date{\today}

\begin{abstract}
Motivated by a recent measurement of proton-proton elastic scattering 
observables up to 3.0~GeV, we investigate the description of those data within 
models of the nucleon-nucleon ($NN$) interaction valid above the pion 
production threshold. 
In addition to including the well known Delta resonance we incorporate
two low-lying $N^\ast$ resonances, the $N^\ast$(1440) and the $N^\ast$(1535), 
and study their influence on $pp$ and $np$ observables for projectile 
laboratory kinetic energies up to 1.5~GeV.  
\end{abstract}


\maketitle



\section{Introduction}
\label{sec-intro}

Elastic nucleon-nucleon ($NN$) scattering is a process fundamental to our
understanding of the nuclear forces. 
This knowledge forms the basis for a broad range of applications
from few-nucleon reactions to heavy ion physics.  
Consequently, many experimental and theoretical studies have been devoted to 
the subject. 
Since the measurements at, e.g., LAMPF \cite{mcnaughton,Arndt1}, the database 
has doubled, and global phase shift analyses now extend to 3.0~GeV projectile 
kinetic energy for $pp$ scattering and 1.3~GeV for $np$ 
scattering~\cite{Arndt:2007qn,SAID}.
Measurements of polarization observables (analyzing power, spin correlation 
coefficients, depolarization parameters and spin transfer triple coefficients)
at discrete kinetic energies up to 2.9~GeV were obtained by the NNSaturne II 
Collaboration~\cite{Ball00,allgower}.
At lower kinetic energies (0.2 - 0.45~GeV), precise analyzing powers and spin 
correlation coefficients have been provided by the PINTEX 
Collaboration~\cite{Haeberli97,Prz98,Rat98} using the IUCF Cooler Ring 
and an internal polarized hydrogen target.
At higher kinetic energies (0.45 - 2.5~GeV), precise and internally consistent 
excitation functions, i.e., differential cross sections
\cite{Albers97,Albers04}, analyzing powers \cite{Altmeier00,Altmeier05} and 
spin correlation parameters \cite{Bauer03,Bauer05}, have recently been 
measured by the EDDA Collaboration using the Cooler Synchrotron COSY and 
internal targets. 
These new proton-proton elastic scattering data are included in the new global
phase shift analysis up to 3.0 GeV kinetic energy with its most recent
solution SP07~\cite{Arndt:2007qn}.

The theoretical description of the nuclear force also has quite a long history.
During the last decade a lot of effort has been directed toward deriving the 
nuclear force in the framework of effective field theory (EFT), especially in
the form of chiral perturbation theory ($\chi$PT).
This EFT can be regarded as the low-energy limit of the underlying theory
quantum chromodynamics (QCD) and inherits its spontaneously broken chiral
symmetry.
As first pointed out by Weinberg~\cite{We90, We91}, the nucleon-nucleon 
potential can be expanded in powers of the $\chi$PT expansion parameter 
$P \equiv\frac{p,m_\pi}{\Lambda_\chi}$, where $p\sim m_\pi$ is a typical  
nucleon three-momentum and $\Lambda_\chi\sim1$~GeV is the chiral scale. 
Weinberg's idea has since been extended to several orders in the
chiral expansion with $NN$ potentials of $\chi$PT derived to NNLO
in Refs.~\cite{Or96,Ka97,Ep99} and to N$^3$LO~\cite{EM03,Ep05}. 
At N$^3$LO the description of $NN$ data below $E_{\rm lab}=200$ MeV is 
comparable in quality to that obtained in the so--called 
``high-precision'' $NN$ potential models \cite{av18,nijm,cdbonn}, 
provided cutoffs in the range 500--600 MeV are considered~\cite{EM03}.
However, the energy range appropriate for the $\chi$PT expansion is determined 
by the expansion parameter $P$ being ``small''.
Thus energy regimes of $\sim$~0.5~GeV and higher are not accessible by
currently employed EFT methods. 

In the later 1980's and early 1990's considerable theoretical effort was
directed towards understanding the $NN$ interaction in the energy region
between 0.4 and 1.0~GeV, which is dominated by the Delta isobar. 
These approaches may roughly be divided into two classes \cite{cefb12}. 
Firstly, there are the models based on a formulation of  genuine unitary 
($NN\pi$) three-particle equations, explicitly including the $\pi N$ scattering
amplitudes $P_{33}$ [Delta ($\Delta$) resonance] and $P_{11}$ (Roper resonance)
and usually employing separable potentials.
The second approach is based on the extension of conventional low-energy 
potentials into the single-pion production 
region~\cite{hlee,Faassen1,Faassen2,bonnce}. 
Most of these are based on coupled channel calculations containing $NN$, 
$N\Delta$ and $\Delta\Delta$ channels, and the single pion production is 
incorporated through the decay of the Delta isobar doorway states.  
The main differences between the two types of models are the choice of 
kinematics, the description of the width of the Delta resonance, and the choice
of $NN$ potential being extrapolated.

More recent theoretical work on the $NN$ interaction in the energy domain
around 0.5~GeV and higher was driven by the need of a better understanding of
pion production in different reactions, e.g., in pion photoproduction from the 
deuteron in the Delta region \cite{schwamb,sato} and pion production in 
heavy-ion reactions \cite{HA94}, measured at GSI. 
The exploratory study of Ref.~\cite{Eyser} investigates how well a meson 
exchange model containing the Delta resonance together with a Wood-Saxon type 
optical potential can describe cross sections and polarization 
observables in the region between 1 and 3~GeV. 
The authors point out that the polarization observables in $pp$ scattering for 
energies above 1~GeV are badly represented by the model employed.  

In this work we want to concentrate on the structure of the nuclear force as 
revealed in $NN$ scattering between 1 and 1.5~GeV 
projectile kinetic energy, the so-called 2nd resonance region. 
This regime is dominated by low lying, well-isolated nucleon resonances. 
A better understanding of the properties and parameters of 
those resonances is 
now being pursued at Excited Baryon Analysis Center 
(EBAC)~\cite{JuliaDiaz:2007kz} based on 
photoproduction data at ELSA (Bonn), JLab (Newport News), 
MAMI (Mainz), GRAAL (Grenoble), and Spring-8 (Hyogo).  
We want to study the influence of the lowest lying
$\pi N$ resonances on the $NN$ data in the region around 1~GeV,  
which is slightly beyond the region dominated by the Delta resonance. 
We start from a meson exchange model and introduce the lowest order 
loop diagrams containing the $\Delta$, Roper, and $S_{11}$ resonances into the
model. In Section~II we give the underlying interaction Lagrangians and in
Section~III we describe the resonance part of the potential and
the parameterization of their width. In Section IV we describe our data fitting
procedure, and in Section V we discuss our results obtained when successively 
introducing the resonances into the potential model. 
We summarize and conclude in Section VI.


\section{Relativistic Meson-Exchange model for \boldmath$NN$ 
scattering above the pion threshold}
\label{sec-lagrangians} 

For projectile energies in the GeV range it is mandatory to employ a 
relativistic formulation of $NN$ scattering. 
A suitable starting point would be the four-dimensional Bethe-Salpeter 
equation~\cite{BS}, which however is very difficult to solve. 
The only successful solution for $NN$ scattering including Delta isobar 
($\Delta$) degrees of freedom was obtained in Ref.~\cite{Faassen1}. 
Since three-dimensional scattering equations are more amenable for 
computations, so-called three-dimensional reductions of 
the Bethe-Salpeter equation have been used~\cite{Faassen2}. 
In this work we will use the reduction proposed by Thompson~\cite{Thompson},
which was also used in Ref.~\cite{Eyser}. 
For completeness we give the invariant scattering amplitude ${\hat T}$ as 
introduced by the Thompson reduction:
\begin{equation}
  {\hat T}({\bf p}',{\bf p},W) = {\hat V}({\bf p}',{\bf p}) + 
  \int d{\bf k} \; {\hat V}({\bf p}',{\bf k}) \frac{M^2}{E^2_k} \frac{1}{W
    +i\varepsilon -2E_k} {\hat T}({\bf k},{\bf p},W).
  \label{eq:2.1}
\end{equation} 
Here $W$ is the invariant mass of the $NN$ system, $W=2E_p$, and
$E_p=\sqrt{p^2+M^2}$ in the c.m.\ system. 
Redefining the scattering amplitude as well as the
potential as
\begin{eqnarray}
  T({\bf p}',{\bf p},W) & = & \frac{M}{E_{p'}} {\hat T}({\bf p}',{\bf p},W)
  \frac{M}{E_{p}}, \\
  V({\bf p}',{\bf p}) & = & \frac{M}{E_{p'}} {\hat V}({\bf
    p}',{\bf p}) \frac{M}{E_{p}}, 
\label{eq:2.2}
\end{eqnarray} 
leads to a scattering equation which has the form of a standard
Lippmann-Schwinger equation with relativistic kinetic energies
\begin{equation}
  T({\bf p}',{\bf p},W) = V({\bf p}',{\bf p}) + \int d{\bf k} \;  V({\bf p}',
  {\bf k}) \frac{1}{W +i\varepsilon -2E_k} T({\bf k},{\bf p},W),
\label{eq:2.3}
\end{equation}
which can be solved with standard numerical techniques.

The pseudoscalar pion provides the long-range part
of the nuclear force and most of the tensor forces. 
Shorter range contributions of the potential are being represented by the 
exchange of heavier mesons, the vector mesons $\rho$ and $\omega$, representing
the $2\pi$ $P$-wave resonance and a $3\pi$ resonance. 
The $\omega$ provides most of the short range repulsion of the nuclear force. 
The intermediate attraction is provided by the exchange of a light 
scalar-isoscalar ``$\sigma$"-meson, which has not been directly observed. 
A recent reanalysis \cite{leutwyler} of $\pi\pi$ data gave evidence for the 
existence of a very broad resonance at threshold with the quantum numbers of 
the vacuum, a mass of about 440~MeV, and a width of about 272~MeV. 
For a simple model which still can capture basic features of the $NN$ 
interaction the exchange of $\pi$ (138~MeV), ``$\sigma$" ($\simeq$ 500~MeV), 
$\rho$ (769~MeV), and $\omega$ (783~MeV) is sufficient.
The coupling of these mesons to the nucleon is given by the following 
interaction Lagrangians
\begin{eqnarray}
  {\cal L}_{pv} & = & 
  -\frac{f_{ps}}{m_{ps}} {\bar \psi}\gamma^5 \gamma^\mu\psi 
  \partial_\mu\varphi^{(ps)},\\
  {\cal L}_{s} &=& - g_s {\bar \psi} \psi \varphi^{(s)}, \\
  {\cal L}_{v} &=& -g_v {\bar \psi} \gamma^\mu \psi \varphi^{(v)}_\nu -
  \frac{f_v}{4M} {\bar \psi}\sigma^{\mu\nu}\psi(\partial_\mu \varphi^{(v)}_\nu
  -\partial_\nu \varphi^{(v)}_\mu).
\label{eq:LOBE}
\end{eqnarray}
Here $M$ is the nucleon mass, and $m_\alpha$, with $\alpha = pv, ps, s, v$,
denotes the masses of mesons with pseudo-vector, pseudo-scalar, scalar and
vector character. 
If the mesons carry isospin, the field $\varphi^{(\alpha)}$ should be replaced 
by ${\bm\tau} \cdot \bm\varphi^{(\alpha)}$, where $\tau^i$ are the usual 
Pauli spin matrices. 
Very often one-boson-exchange (OBE) models also include the $\eta$
(547 MeV) and the $a_0$ (980 MeV) \cite{bonnphysrep}.  
However, if the scalar-isovector $a_0$ is included, there is no reason to not
also consider the scalar-isoscalar $f_0$ (980 MeV) with similar mass, 
as well as the $\eta'$ (958 MeV).  
It is clear that mesons with large masses will contribute to the short 
range part of the $NN$ interaction. 
Thus their effect is very small when considering low energies, and there 
they can in principle be omitted. 
However, since we want to investigate a meson exchange model at laboratory 
energies around 1 GeV and higher, we include for consistency all 
strangeness-zero mesons with masses below 1~GeV. 

\begin{figure}
\begin{center}
\includegraphics[width=10cm]{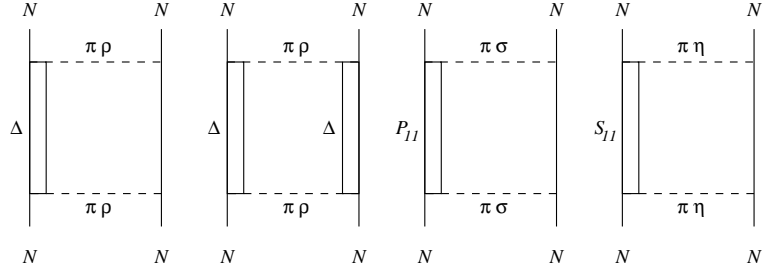}
\end{center}
\caption{The two-meson exchange iterative contributions involving nucleons
(represented by solid lines) and nucleon resonances (double line). The dashed
lines represent the exchanged mesons. The nucleon resonances and exchanged 
mesons are specified in the figure.
\label{fig1}}
\end{figure}

In order to introduce contributions from the nucleon resonances into the 
potential model, we consider the lowest order loop diagrams containing a 
resonance as an intermediate state. 
The first diagrams to consider are those with a nucleon and a single 
resonance ($\Delta$, $P_{11}$, $S_{11}$) in the intermediate state. As a first
step we want to consider here only low lying $S$- and $P$-wave resonances.  
These resonances can be on-shell  at the $NN$ 
laboratory kinetic energies $T_{\rm lab}=632$~MeV for the $N\Delta$, 1136~MeV for the
$NP_{11}$, and 1381~MeV for the $NS_{11}$ diagrams, respectively. 
These energies are given in Table~\ref{table-1} together with the pion and eta
production thresholds.
The energy at which two Delta's are on shell in an intermediate state is  
below the threshold for the $NS_{11}$ intermediate state,
therefore the diagram with $\Delta\Delta$  
intermediate states is also included in our calculation. The diagrams included
are shown in Fig.~\ref{fig1}.
 
Since the Delta is an isospin-3/2 state, only isovector mesons can excite it. 
The only possible exchange mesons are then $\pi$ and $\rho$.
Note that in order to fulfill the weak unitarity bound 
($\sigma_{\rm tot} > \sigma_{\rm el}$), diagrams with $\pi$-$\pi$, 
$\pi$-$\rho$, and $\rho$-$\rho$ exchange must all be included. 
The $P_{11}$ resonance has the quantum numbers of the
nucleon and decays predominantly into $N\pi$ (60-70\%) \cite{PDG}
and to a lesser extent to $N\pi\pi$. 
We consider here the exchange of a pion and in addition the exchange of a
``$\sigma$'', which mimics (correlated) two-pion exchanges. 
The decay of the $P_{11}$ into $N\rho$ has a probability of less than 8\% and 
thus this $\rho$ exchange is not considered.
The $S_{11}$ resonance is the only known resonance that couples strongly to the
$N\eta$ channels. 
The dominant decay channels are $N\pi$ and $N\eta$~\cite{PDG}.
Thus, for this resonance we consider only the exchange of $\pi$ and $\eta$.
Again, in order to fulfill the weak unitarity bound, each of the diagrams in
Fig.~\ref{fig1} represents in reality four separate diagrams.

The interaction Lagrangians for the couplings to the Delta are given by
\begin{eqnarray}
  {\cal L}_{N\Delta\pi} &=& -\frac{f_{N\Delta\pi}}{m_\pi} {\bar \psi} {\bf T} 
  \psi^\mu\partial_\mu {\mathbf \varphi}^{(\pi)} + {\rm H.c}. \\
      {\cal L}_{N\Delta\rho} &=& -i\frac{f_{N\Delta\rho}}{m_\rho}  
      {\bar \psi} \gamma^5 \gamma^\mu {\bf T} \psi^\nu 
      \left( \partial_\mu {\mathbf \varphi}^{(\rho)}_\nu -
      \partial_\nu {\mathbf \varphi}^{(\rho)}_\mu \right) +{\rm H.c.},
\label{eq:Ldelta}
\end{eqnarray}
where $\psi_\mu$ is the Rarita-Schwinger field \cite{rarita,lurie,dumbrajs}
describing the spin-3/2 Delta-isobar and $\psi$ stands for the nucleon field.
The operator ${\bf T}$ acts between isospin-1/2 and isospin-3/2 states and
H.c.\ denotes the Hermitian conjugate.

The $P_{11}$ resonance has the quantum numbers of the nucleon with the 
interaction Lagrangians given by
\begin{eqnarray}
  {\cal L}_{NN^*_{P_{11}}\pi} &=& -\frac{f_{NN^*_{P_{11}}\pi}}{m_\pi} \; 
{\bar \psi}_{P_{11}} \gamma^5 {\bf \tau} \psi \partial_\mu 
 {\mathbf \varphi}_\pi + {\rm H.c.}, \\
  {\cal L}_{NN^*_{P_{11}}\sigma} &=& -g_{NN^*_{P_{11}}\sigma} \;  
  {\bar \psi}_{P_{11}} \psi \varphi_\sigma + {\rm H.c.}
\label{eq:L1440}
\end{eqnarray}
In contrast to the $P_{11}$ resonance, the $S_{11}$ resonance has negative 
parity, and the interaction Lagrangians are given by \cite{Benmerrouche}
\begin{eqnarray}
  {\cal L}_{NN^*_{S_{11}}\pi} & = & 
  -i g_{NN^*_{S_{11}}\pi} \; {\bar \psi}_{S_{11}}
   {\mathbf \tau} \; \psi {\mathbf \varphi}_\pi+{\rm H.c.}, 
   \label{eq:L1535pi} \\
        {\cal L}_{NN^*_{S_{11}}\eta} &=&  -i g_{NN^*_{S_{11}}\eta} \; 
        {\bar \psi}_{S_{11}} \psi \varphi_\eta + {\rm H.c.}
\label{eq:L1535}
\end{eqnarray}

All the meson-baryon vertices described by the above Lagrangian
interactions are modified by form factors of the dipole-type
\begin{equation}
  F_\alpha \left( ({\bf p}'-{\bf p})^2\right) = \left( \frac{\Lambda_\alpha^2 -
    m_\alpha^2}{\Lambda_\alpha^2 +({\bf p}'-{\bf p})^2} \right)^{n_\alpha},
\label{eq:dipolecut}
\end{equation}
where $({\bf p}' -{\bf p})$ is the momentum transfer between the two 
interacting baryons. 
The parameter $\Lambda_\alpha$ is the so-called cutoff mass and is
determined by fitting $NN$ data. 
We choose $n_\alpha =1$ for all vertices with the exception of the 
$N\Delta\rho$ vertex, where $n_\alpha =2$ is applied.

\section{Resonance Contributions to the Potential}
\label{sec-resonances}

The nucleon resonances enter the $NN$ potential via the loop diagrams given in
Fig.~\ref{fig1}. 
Schematically, the contribution to the potential 
from these ``iterative'' diagrams can be written as
\begin{equation}
  V_{NN,R_1R_2} \; \frac{1}{W-E_{R_1} - E_{R_2}} \; V_{R_1R_2,NN}.
  \label{eq:3.1}
\end{equation}
Here $W$ is the invariant mass of the $NN$ system, and $E_{R_1}$, $E_{R_2}$
the energies $E_{R_i}=\sqrt{m^2_{R_i}+k^2}$ of the intermediate states, which 
can be either a nucleon, a $\Delta(1232)$, a $N^*(1440)$ or a $N^*(1535)$. 
The transition potentials $V^\pi_{NN,N\Delta}$ and $V^\pi_{NN,\Delta\Delta}$ 
can be found in Ref.~\cite{HMpi} and $V^\rho_{NN,N\Delta}$ and
$V^\rho_{NN,\Delta\Delta}$ in Ref.~\cite{HMrho}. 
The transition potentials for the two $N^*$ resonances can be calculated in a 
straightforward fashion, since they carry isospin 1/2. Since the resonances 
have a finite life time, their mass is modified by the width of the resonance 
according to 
\begin{equation}
  \mu_{r_i} = m_{r_i} - i\frac{\Gamma_{r_i}}{2}.
  \label{eq:3.2}
\end{equation}
Because of the finite width $\Gamma_{r_i}$, the mass, and thus the energy of 
the resonances, acquires an imaginary contribution.
The loop diagrams of Fig.~\ref{fig1} then contain an imaginary piece, which 
describes the inelasticity of the $NN$ scattering process above the
pion threshold.
The effect of direct pion production in $NN$ scattering up to 1~GeV has been 
studied in Refs.~\cite{Faassen1,bonnunitary}, and found to be very small in 
comparison to pion production via, e.g., the $\Delta$-resonance. 
For the present purpose $\pi$ (and $\eta$) production will be restricted to 
occur only via resonance doorway states.

We follow here the same parameterization of the penetrability as given in
Ref.~\cite{HA94}, i.e.,
\begin{equation}
  \Gamma = \gamma q v_L(qR). 
  \label{eq:3.3}
\end{equation}
Here the momentum $q$ is the c.m.\ momentum of the $\pi N$ system [or $\eta N$
system in case of the $N^*(1535)$], and should not be confused with the
momenta in the $NN$ system. 
The parameter $R$ is fitted and chosen such that the product $qR$ becomes 
dimensionless. 
We choose $R=6.3$~GeV$^{-1}$ for all three resonances under consideration. 
The dimensionless quantity $\gamma$ is the reduced width, and the penetrability
factor $v_L$ is defined as
\begin{equation}
  v_L(qR) = \frac{1}{(qR)^2 n^2_L(qR) + (qR)^2 j^2_L(qR)},
  \label{eq:3.4}
\end{equation}
where $L$ denotes the orbital angular momentum quantum number of the resonance.
The functions $j_L$ and $n_L$ are the spherical Bessel and Neumann functions.
Thus for the resonances $\Delta$ and $N^*(1440)$ with the quantum numbers of 
$P_{33}$ and $P_{11}$ the width is given as
\begin{equation}
  \Gamma_{\Delta,P_{11}} = \gamma_{\Delta,P_{11}} \frac{q^3 R^2}{1+q^2R^2} \; \Theta
(q-q_{thr}) 
  \label{eq:widthdel}
\end{equation}
and for the resonance $N^*(1535)$ with the quantum numbers $S_{11}$ 
\begin{equation}
  \Gamma_{S_{11}}= \gamma_{S_{11}} q \; \Theta (q-q_{thr}),
  \label{eq:widthS11}
\end{equation} 
where the step function $ \Theta (q-q_{thr})$ ensures that the pion (eta) 
production starts at the corresponding thresholds in the $NN$ c.m.
frame.
The expressions for $\Gamma_{r_{i}}$ show that the resonance width increases linearly
with increasing momentum $q$ and introduces a linear divergence in the 
imaginary part of the potential. 
Since this does not seem physical plausible, Ref.~\cite{HA94} introduced an 
additional phenomenological form factor to the resonance width of the form
\begin{equation}
  Z(q^2) = \frac{q^2(m^2_{r_i}) +\kappa^2}{q^2(s) +\kappa^2},
  \label{eq:3.7}
\end{equation}
so that Eq.~(\ref{eq:3.3}) is modified to
\begin{equation}
  \Gamma = \gamma q v_L(qR) \; Z(q^2).
  \label{eq:3.8}
\end{equation}
As already mentioned in Ref.~\cite{HA94}, the analytic form of the form factor 
is relatively unimportant, as long as the resonance curve is qualitatively
described, i.e., it peaks at the resonance position. 
However, at the energies we are considering, setting $Z(q^2)=1$ has only 
minor influence on the numerical calculation of the $NN$ observables.
Even so, we choose to employ Eq.~(\ref{eq:3.7}) for the physical reason
already mentioned.
The form factor is defined such that it is $1$ at the exact resonance position,
$s=m^2_{r_i}$, of the meson-nucleon system. 
The parameters $\gamma_{r_i}$ are fitted such that the resonance width
at the resonance position coincides with the experimentally extracted 
width~\cite{PDG}. 
A similar behavior has been observed in Ref.~\cite{bonnce} where the resonance
width is calculated as the imaginary part of the lowest order $\Delta N\pi$
loop diagram. There a dipole cutoff with the same cutoff parameter $\Lambda_{N\Delta\pi}$ 
as employed in the $NN$ potential was used. 

Since we consider only single $\pi$ (or $\eta$) production, we fit the 
width $\Gamma_{r_i}$ to the corresponding partial widths as given in 
Ref.~\cite{PDG}.
However, the $\Delta$ resonance decays with a probability $>$~99\% into $N\pi$.
Thus, in this case the partial width is equal to the total width, which lies 
between 115 and 125~MeV. 
For the $N^*(1440)$ resonance we use the partial width
for the $N\pi$ decay to determine the strength parameters. 
The total width has a wide ranging from 200~MeV 
to 450~MeV Ref.~\cite{PDG}.
We choose the value
$\Gamma^{\rm tot}_{P_{11}}$~=~391~MeV, from Ref.~\cite{Manley} to determine  
the partial width.

The $N^*(1535)$ resonance decays mostly into $N\pi$ and $N\eta$. 
Up to the $\eta$-production threshold, which corresponds to the $NN$
laboratory energy 1253~MeV, we only consider the partial width for $N\pi$ 
decay, above the $\eta$ threshold we add also the partial width for the 
$N\eta$ decay. 
The parameters we use for describing the width of the resonances included in 
our model are given in Table~\ref{table-2}. 
We tested the influence of the width cutoff parameter $\kappa$ on the $NN$ 
observables.
Varying $\kappa$ between about 200 and 1000~MeV did not lead to any visible 
difference in their description. 
Thus, for the $\Delta$ resonance we used the value (200 MeV) given in 
Ref.~\cite{HA94}, while we for the two $N^*$ resonances we adopted an average 
value of 400~MeV.

\section{Fitting procedure}
\label{sec-fitting}

The parameters in our model are the masses of the mesons and baryons, together
with the coupling constant and cutoff parameters for the baryon-baryon-meson 
vertices.
With one exception the masses are fixed to the values recommended by the 
Particle Data Group (PDG)~\cite{PDG}. 
For the nucleon mass we employ the average of the neutron and proton mass, 
$M=938.926$~MeV.
Similarly, for the pion mass we use the average of the charged and 
neutral pion masses. 
The only mass parameters which are not fixed are those of the ``$\sigma$'' 
mesons (one for each isospin channel), which we consider as free parameters. 
Since most meson-baryon coupling constants are poorly known, we treat them 
as free parameters, which are allowed to vary within a certain range. 
We use the values of Refs.~\cite{bonnce,D52} as starting points. 
The only exception is the $NN\pi$ coupling, which we fixed to
$g^2_{NN\pi}/4\pi=13.8$, according to recent measurements at 
IUCF~\cite{Vigdor04} and extractions by the Nijmegen group~\cite{Stoks:1992ja}.
All cutoff parameters $\Lambda$  are treated as free parameters
and we allowed them to vary between 0.8 and 2.5 GeV.  
A more detailed discussion of the parameters obtained in the different fits 
will be given in the next section.

Our goal is to find parameter sets for which our models give a good description
of the $NN$ observables, i.e., total and differential cross sections, as well 
as spin observables, up to 1.5~GeV. 
However, we do not fit directly to data but rather to a partial wave analysis,
since in some cases the amount of data is rather sparse. 
As reference values we choose the SAID energy dependent phase shift solution 
Sp07 from the CNS Data Analysis Center \cite{cns}. 
However, we need to prepare from this solution a ``data set" that is suitable 
for our fitting procedure. 
First, since the imaginary parts of our model are already fixed by fitting the 
widths of the resonances to their experimental values (see 
Section~\ref{sec-resonances}), the partial wave inelasticities $\rho_l$ 
are already given. 
Therefore, we need to restrict ourselves to fitting only the phase shifts
$\delta_l$. 
Also, since we keep the pion coupling constant $g^2_{NN\pi}$ fixed, we should 
not include higher partial waves in the fit. 
What constitutes a high partial wave is expected to depend on the energy 
considered.
Therefore, we divide the energy range up to 1.5~GeV into three separate 
regions: up to 0.3~GeV laboratory kinetic energy we include only partial waves 
$J \leq 2$ as reference values, between 0.3 and 0.8~GeV we consider all 
partial waves with $J \leq 4$, and above 0.8~GeV we include in addition all 
partial waves with $J \leq 6$. 
However, since there are only very few experimental data points for $np$ 
scattering above 0.8~GeV, we exclude isospin-0 partial waves from the 
SAID data base above 0.8~GeV. 

In addition we need to account for the fact that the SAID energy-dependent
solutions do not give errors.
The absolute errors for each partial wave are estimated with a simple
polynomial reflecting the errors of the single energy solution.
In this way we try to account for the accuracy of the phase shift analysis,
which varies with energy and partial waves due to different amount of
available data.
Especially partial waves with small absolute values get a very large relative
error assigned, since they would be overrepresented
if only simple relative errors were used.
However, in order to avoid obscuring the plots, these errors are not
included in our figures.
In all cases, the assigned errors increase monotonically with energy and
reach a maximum absolute value of 2 at 1.5 GeV. We tested different error
setups and choose the one giving the best overall results. All partial waves
get similar errors assigned. However additional weight-factors could be
included during the fitting procedure itself, e.g. the $^3$P$_2$ partial wave
above 800~MeV was assigned a low weight in the final fits. 
After creating our ``data base'' according to these specifications, 
we perform a nonlinear least-square fit using the Minuit program provided by 
the CERN library~\cite{Minuit}. 
A discussion of the resulting model parameters and the description of the
phase shifts and observables will be given in the following section. 

The resonances are added one-by-one to our model, starting form a traditional 
meson-exchange $NN$ scattering model.
This way the influence of the different resonances can be understood separately
from each other.

\section{Results and Discussion}
\label{sec-results}

\subsection{Including the \boldmath$\Delta$(1232) -- $P_{33}$  Resonance}
\label{sec-results-delta}

As a first step we consider only a meson exchange model together with the 
excitations of the $\Delta$(1232) resonance.
This is similar to the models used in Ref.~\cite{D52,Eyser}. 
Here our focus is to create an updated model to which we can later add the 
$P_{11}$ and $S_{11}$ resonances.  
Thus,  we first concentrate on the description of $NN$ data up to 1~GeV, the 
energy regime which is expected to be dominated by the $\Delta$(1232) 
resonance. 
In addition we include the contributions of the mesons $a_0$, $f_0$, and 
$\eta '$, which all have a mass slightly below 1~GeV and were not included in 
\cite{D52,Eyser}. 
The influence these mesons have on the description of the phase shifts is 
relatively small and the coupling constants are determined to be of the order 
unity by our fitting procedure. 
They were allowed to vary within the interval [0.0,5.0] and the fit always 
stayed within these boundaries.

The $\rho$-meson tensor-to-vector coupling ratio was allowed to vary within the
limits of 5.9 and 6.3.
The fit prefers a value of 5.9, to which the ratio was kept fixed in the 
subsequent investigations. 
In the case of the $\omega$ meson this ratio is around -0.12 and of negligibly 
small influence. We thus fixed it to zero. 
We also studied the interplay between $\eta$ and $\eta '$ by 
alternately fixing one and letting the other one float. 
It turns out that, for the description of the $NN$ phase shifts, only the 
combined strength of the two seem to matter, i.e., if one is fixed larger (on 
the order of 5) the other one will go to a smaller value, closer to 1.
Therefore we always choose similar starting values for both of them, and let 
them vary between 1 and 6. 
Both couplings stayed within this range. 
Quark model prediction \cite{Downum:2006re} give slightly higher values, but 
do predict both of them to be of the same magnitude. 
For the ``$\sigma$'' meson we let the coupling constants, masses, and cutoffs 
vary separately in the $T=1$ and $T=0$ channels.
However, the values for the masses always stayed relatively close together. 

Finally, we investigated the sensitivity of the  $\Delta$(1232) couplings to 
the overall description of the phase shifts. 
In \cite{D52,Eyser} the  value of $f^2_{N\Delta\pi}/4\pi =$~0.35 is used. 
Similarly, the $\pi N$ model from Ref.~\cite{Schutz:1998jx} employs 
$f^2_{N\Delta\pi}/4\pi =$~0.36. 
On the other hand, quark models relating $f^2_{N\Delta\pi}$ to 
$f^2_{NN\Delta\pi}$ lead to values as low as 0.22 \cite{Brown:1975di}. 
Because of this discrepancy we let the $f^2_{N\Delta\pi}$ float between those 
two values to see if the phase shifts show a preference. 
In addition we also let the cutoff $\Lambda_{N\Delta\pi}$ vary. 
The fit always prefers the lower value of the coupling constant and goes 
to the lower limit of the cutoff boundary, which was set for all couplings 
to 800~MeV.  
After lowering that bound even further, the cutoff $\Lambda_{N\Delta\pi}$ 
stabilizes at values between 600 and 700~MeV. 
Our fit runs consistently prefer a small value of $f^2_{N\Delta\pi}$, so
we fixed this coupling for all further studies to 0.224. 
Since we independently fitted the width of the $\Delta$ resonance to its 
experimental value, this choice of coupling constant does not influence the 
size of the inelasticity provided by the $N\Delta\pi$ contribution. 
A similar investigation of the $f^2_{N\Delta\rho}$ coupling did not reveal 
any sensitivity with respect to the $NN$ phase shifts. 
Thus we kept the value for $f^2_{N\Delta\rho}/4\pi$ the same as in
Refs.~\cite{D52,Eyser} and only varied the cutoff parameter 
$\Lambda_{N\Delta\rho}$.

In total we varied 22 parameters in this model, which includes the 
$\Delta$(1232) resonance as the only means for pion production. 
It is well known that models of this kind are able to describe $NN$ scattering 
up to about 1~GeV in semi-quantitative terms.  
A good parameter set obtained from our fit procedure is given as model (A) in 
Table~\ref{table-3}. 
Using those parameters we calculated $pp$ and $np$ observables as function of 
the projectile laboratory kinetic energy. 
The total cross section for elastic and inelastic $pp$ scattering are shown 
in Fig.~\ref{fig3} as a long-dashed line together with the SAID Sp07 
analysis~\cite{SAID}. 
As already pointed out in Ref.~\cite{Eyser}, the inelastic contributions of a 
model of this type are not large enough to account for the experimentally 
observed ones.
Consequently the total elastic cross section is being overpredicted beyond 
1~GeV. 
This tendency can also be observed in the differential cross section 
for $pp$ scattering in Fig.~\ref{fig4}, especially at the larger angles. 
A partial remedy would be double pion production, which starts to 
contribute above 590~MeV.
We will not attempt to include such amplitudes in the present paper.
The differential cross section for $np$ scattering (Fig.~\ref{fig5}) is 
experimentally less well determined than for $pp$ and data above 800~MeV are 
sparse. 
Overprediction of the $np$ differential cross section is only visible at 
the largest c.m.\ angle.  

The analyzing power $A_N$ is shown in Fig.~\ref{fig6} for $pp$ and in 
Fig.~\ref{fig7} for $np$ scattering.
Above about 800~MeV the analyzing power for $pp$ scattering is strongly 
overpredicted, especially for the larger angles. 
This phenomenon was already pointed out in Ref.~\cite{Eyser}. 
The $np$ analyzing power is also overpredicted at similar angles, however at 
angles larger than 90$^\circ$ our calculation describes the existing data 
quite well.
The spin correlation coefficients $A_{SS}$, $A_{NN}$, and $A_{LL}$
for different c.m.\ angles are shown as function of projectile kinetic energy 
in Figs.~\ref{fig8}--\ref{fig10} for $pp$ scattering and in 
Fig.~\ref{fig11} for $np$ scattering.
For the spin-correlation coefficient $A_{SS}$ the model predictions deviate from 
the data in a similar way as for the analyzing power, whereas the other 
spin-correlation coefficients are described reasonably well. 

In addition to the energy dependence of observables (for fixed angles) we 
also select three different fixed energies for c.m.\ angular distributions of
$pp$ observables.
Those energies are 400~MeV (Fig.~\ref{fig12}), 800~MeV (Fig.~\ref{fig13}), and 
1300~MeV (Fig.~\ref{fig14}). 
As can be expected from the previous discussion, at 400~MeV our calculations 
are in reasonable agreement with the data, while deviations start to show up 
at 800~MeV and become more severe at 1300~MeV. 

The  partial wave phase shifts of our model prediction (A) up 
to J=4 are shown as long-dashed 
lines in Fig.~\ref{fig15} ($T=1$) and Fig.~\ref{fig16} ($T=0$). 
The imaginary parts of the partial wave  amplitudes are given in 
Fig.~\ref{fig17} ($T=1$) and Fig.~\ref{fig18} ($T=0$).

\subsection{Including the \boldmath$N^\ast(1440)$ -- P\boldmath$_{11}$ 
Resonance}

The $N^\ast(1440)$ (Roper) resonance is the lowest $J=\frac{1}{2}$ nucleon 
excitation and thus the first nucleon resonance to add.
Since it has the same quantum numbers as the nucleon this resonance couples 
to exactly the same meson fields as the nucleon, though not necessarily with
the same strength.
We are using the decay modes of the Roper as guide to which meson exchanges to
include. 
It decays predominantly into the $N\pi$ channel (60-70\%)~\cite{PDG}. 
A very crude estimate from the width of the Roper resonance indicates that
the $NN^\ast\pi$ coupling is about 40\% of the $NN\pi$ coupling.
In Ref.~\cite{Riska:2000gd}, Riska and Brown extract an average 
$f^2_{NN^*\pi}/4\pi=0.0012$ from the width and derive an even smaller quark 
model estimate.
On the other hand, a coupled channel model for $\pi N$ 
scattering~\cite{Schutz:1998jx} uses $f^2_{NN^*\pi}/4\pi=0.0024$. 
We choose a value close to the latter, since the overall effect of the 
iterative $NN^\ast$(1440) diagram is quite small. 
The second largest decay channel of the Roper is $N\Delta\pi$ (20-30\%) and 
PDG gives a value of 5-10\% for $N(\pi\pi)_{{\rm s-wave}}$. 
To mimic the latter effect, at least for the real part of the amplitudes, we 
introduce the ``$\sigma$'' meson, and choose a relatively small value for its
coupling.
During the fitting procedure this value stayed small and none of our fits 
reached the large value favored in Ref.~\cite{Schutz:1998jx}. 
The calculations in Ref.~\cite{HA94} introduced additional diagrams containing
$\rho$ and $\omega$ exchange in the iterative $NN^\ast$(1440) diagram,
though with relatively small coupling constants. 
As a test we included the $\omega$ meson, but since the overall effect
in phase shifts and observables is minimal, we did not pursue the inclusion 
of vector mesons any further. 

Adding the $N^\ast(1440)$ resonance to model (A), without refitting the 
parameters, has only a moderate effect on both the $T=1$ and $T=0$ phase 
shifts. 
However, to obtain a true assessment of the effect, the model parameters need 
to be readjusted.
A qualitative fit to the phase shift including the  $N^\ast(1440)$ resonance
is given by the parameters listed in Table~\ref{table-4}. 
The results are plotted as short dashed  lines in 
Figs.~\ref{fig3}--\ref{fig18}. 

A major effect of the $N^\ast(1440)$ can be seen in the $T=0$ channels, which 
now exhibit pion production at threshold. 
Since $N\Delta$ contributions are forbidden, pion production close to threshold
is provided solely by the decay of the Roper resonance.
Overall, the total cross section for inelastic scattering is slightly 
increased. 
However, the contributions of the $N^\ast(1440)$ are by far too
small to make up the difference with the experimental data. 
Though the changes in the $^1$P$_1$ and $^3$P$_2$ phase shifts are relatively
large in going from model (A) to model (B), the final differences  in the 
observables are much smaller. 

Thus, the inclusion of the Roper resonance has a small overall 
effect in the description of $NN$ scattering. 
However, it is vital for the $T=0$ inelasticities close to threshold.
Similar conclusions were drawn in Ref.~\cite{hlee}. 
We see in our fit that the pion-nucleon cutoff is slightly
reduced when the Roper resonance is included. 
Thus we were tempted to see if we can force $\Lambda_{NN\pi}$ to take a small 
value, e.g., 1~GeV, and still obtain a description of the data of equal
quality by letting the $NN^\ast\pi$ coupling take arbitrary values.
In fact, it has been suggested \cite{holindethomas} that a soft $NN^\ast\pi$ 
form factor can be used in a boson-exchange model provided an additional 
pseudoscalar meson, with a mass slightly above 1~GeV, is introduced. 
We find, however, that despite the Roper resonance having the quantum numbers 
of the nucleon, the iterative $NN^\ast$ diagram with pion exchanges can not 
assume this particular role, and we can not get acceptable fits with a very 
soft pion form factor.

\subsection{Including the \boldmath$N^\ast(1535)$ -- S\boldmath$_{11}$ 
Resonance}

The $N^\ast(1535)$ (S$_{11}$) resonance has the quantum numbers of the nucleon,
except it is parity-odd. 
Thus the functional forms of the vertices have different spin structures, see
Eqs.~(\ref{eq:L1535pi}) and (\ref{eq:L1535}). 
The $N^\ast(1535)$  decays mainly into the $N\pi$ and $N\eta$ channels, and 
thus we include only these mesons when considering the iterative 
diagram with the S$_{11}$ resonance in the intermediate state. 
Values for the coupling $g^2_{NN^\ast \pi}/4\pi$ vary considerably in
the literature. 
The quark model estimate from Riska and Brown \cite{Riska:2000gd} suggests that
it is slightly less then half of the $NN\pi$ coupling, while investigations of 
$\pi N$ scattering give a value as low as 0.001~\cite{Krehl:1999km}. 
A study of $\eta$ photoproduction from the deuteron \cite{Breitmoser:1996dy} 
suggests a value of 0.1. Our value of 0.05 obtained from the fit is well within
these boundaries. 
The $NN$ data are not particularly sensitive to this number as long as it stays
around 0.1. 
For the coupling to the $\eta$ meson we obtained 0.99, which is higher than the
value of 0.33 given in Refs.~\cite{Krehl:1999km,Breitmoser:1996dy}.

When including the $N^\ast(1535)$ resonance (without refitting), its effect is 
most prominent in the $^1$P$_1$ phase shift. 
In general, the $T=0$ phase shifts are more sensitive to the size of its 
contribution. 
However, a true assessment should only be made after a refit of the parameters.
This gives the parameters of Table~\ref{table-5}, which constitute our 
model (C). 
The observables calculated with these parameters are plotted in 
Figs.~\ref{fig3}--\ref{fig14},  partial wave phase shifts in Figs.~\ref{fig15} and 
\ref{fig16}, and the imaginary parts of the scattering amplitudes in 
Figs.~\ref{fig17} and \ref{fig18} (solid lines in all figures). 

In general, the $N^\ast$(1535) resonance has a tendency to counterbalance 
the contributions of the $N^\ast$(1440) resonance.
A description of $NN$ observables with both $N^\ast$ resonances is of about the
same quality as the description containing only the $\Delta$ resonance. 
Only a closer look at the inelasticities in the $np$ phase shifts shows that 
these two resonances are necessary in order to obtain inelastic contributions 
in the region close to threshold.

Comparing the parameters of model (C), given in Table~\ref{table-5}, with the
parameters of the other two models shows that the coupling of the $\omega$ 
meson decreases when introducing the $N^\ast$(1535) resonance, indicating that 
this contribution has an overall repulsive effect. 
The $\rho$ coupling, on the other hand, decreases minimally.  
In addition, the pion cutoff, $\Lambda_{NN\pi}$, assumes a slightly lower
value.

\subsection{Role of the \boldmath$^3P_2$ Partial Wave in the Description of 
Polarization above 800 MeV}

Certain observables are particularly sensitive to specific phase shifts. 
For example, the analyzing power $A_N$ is sensitive to spin-triplet states. 
Above roughly 800 MeV the analyzing power is strongly overpredicted for larger 
c.m.\ angles, as seen in Fig.~\ref{fig6} for the $pp$ excitation function. 
This effect is also seen in the corresponding $np$ analyzing power, however 
not as strongly. 
Thus, one needs to concentrate on the large discrepancies seen in the 
$T=1$ triplet phase shifts. 
The most obvious candidate is the $^3$P$_2$ phase shift, which above 800~MeV 
deviates strongly from the experimental analysis. 
The experimental analysis changes sign at about 1 GeV, while the
model predictions remain positive and roughly constant.

In order to isolate the influence of the $^3$P$_2$ partial wave on the 
observables, especially on the analyzing power, we replaced the phase-shift 
$\delta (^3$P$_2)$ from our model above 800~MeV with the one from the SAID 
analysis and then recalculated the observables. 
Since the inelasticity agrees with the SAID analysis, we 
stayed with our model prediction for this parameter. 
The resulting $pp$ differential cross section and analyzing power are shown as 
function of laboratory kinetic energy for c.m.\ angles 33$^\circ$ and 
53$^\circ$ in Fig.~\ref{fig19}, 
together with the original prediction for comparison. 
The large discrepancy almost completely vanishes. 
In Fig.~\ref{fig20} the differential cross section and analyzing power
at 800 and 1300~MeV are shown as function of c.m.\ scattering angle. 
Again, once the $^3$P$_2$ partial wave is replaced with the SAID value, 
the discrepancy nearly disappears. 
All the other observables are only marginally affected by this change in the
$^3$P$_2$ partial wave.   
The description of the analyzing power for $np$ scattering at 53$^\circ$, 
Fig.~\ref{fig7},  is also improved by the change in the $^3$P$_2$ partial wave,
while the backward angles are not affected and remain reasonably well 
described. 
For higher energies, the $pp$ analyzing power is a smooth function of the 
energy at nearly all angles \cite{SAID}. 
Thus we may speculate that once the fast change of sign 
of the $^3$P$_2$ partial wave between 800 and 1500~MeV wave is explained, the 
$pp$ analyzing power can be described reasonably well.  
However, the present ingredients in our model can not account for a correct 
description of the $^3$P$_2$ partial wave.
The efforts of Ref.~\cite{Eyser} indicate that a variation of model parameters 
close to our model (A) is not able to improve the 
description of the analyzing power as function of energy. 
Instead, a different physical mechanism needs to come into play.
A strong energy dependence in one single partial wave often indicates the 
opening of a new threshold in that particular channel, given by the coupling to
another nucleon resonance. 
Since the $^3$P$_2$ partial wave has a very strong energy dependence above 
800~MeV one may speculate that only a $S$-wave coupling to this partial wave 
can be responsible, which would require a $J=\frac{5}{2}$ resonance.  
The lowest $J=\frac{5}{2}$ resonances have masses around 1.7~GeV and were found to 
be important for a description of $\pi N$ scattering below 2~GeV
\cite{Shklyar:2004dy,JuliaDiaz:2007kz}. Consideration of either the D$_{15}$(1675) or the
F$_{15}$(1680) would be possible candidates to explore this avenue.

\section{Summary and Conclusions}

In this paper we have studied the effects the first few resonances have on the
description of the $NN$ phase shifts in the energy regime around and above 
1~GeV.
We have incorporated the $\Delta$(1232) resonance as well as the $N^\ast$(1440)
and  $N^\ast$(1535) in a relativistic meson-exchange model. 
We started with a model including contributions of only the $\Delta$(1232) 
resonance, which we refitted to $NN$ data up to 1~GeV, and then added the 
contributions of the $N^\ast$(1440) and  $N^\ast$(1535) resonances, now 
extending the fitted region up to 1.5~GeV.
We verified the characteristic deficiencies in describing the $pp$ analyzing 
powers above $\sim$800~MeV that were pointed out already in Ref.~\cite{Eyser}. 
We could isolate this defect as being caused by a poor description of the 
$^3$P$_2$ partial wave above 800~MeV.
This situation is not improved by including also the two $N^\ast$ resonances. 
We speculate that only a new channel, which couples with $L=0$ to the 
$^3$P$_2$ partial wave will be able to cause such a strong energy dependence
in that specific partial wave. 
Further investigations are necessary to investigate this assertion. 

We also find that, in general, the description of available $np$ observables 
seems to be better than that of $pp$ observables. 
However, this finding may be too optimistic due to the small amount
 of $np$ observables above 1~GeV. 
We also observe that the total cross section for inelastic scattering is 
underpredicted by about a factor of two by our models, indicating that not 
enough inelastic channels are considered. 
In Ref.~\cite{Eyser} this was compensated for by introducing a spin-independent
optical potential providing an average absorption. 
As shown there, this provides a reasonable mechanism to incorporate a bulk 
contribution to account for single- and multiple-meson production. 
However, such an optical potential will not be able to account for the strong 
energy dependence in selected partial waves, e.g., the $^3$P$_2$ partial wave 
around 1~GeV. 

In this work the main effort was introducing contributions of two of the lowest $N^\ast$
resonances into the description of $NN$ observables above 1~GeV laboratory kinetic
energy. In order to have uniform description of the widths of the resonances we employ
the relatively simple parameterization suggested in Ref.~\cite{HA94}. This can be improved
by taking advantage of the recent work of the Excited Baryon Analysis Center
(EBAC)~\cite{JuliaDiaz:2007kz}
analyzing $\pi N$ scattering in region below the 2~GeV nucleon resonance region, and e.g.
employing vertex functions and self-energies determined in the $\pi N$ analysis in the $NN$
system. This will allow to make connections between analyses of electromagnetic meson
production reactions and the understanding of the $NN$ system in this energy regime.

\section*{Acknowledgments}
This work was performed in part under the
auspices of the U.~S. Department of Energy under contract
No.~DE-FG02-93ER40756 with Ohio University and the German BMBF under contract
No.~06BN664I(6).
It was also supported by the U.~S. National Science Foundation 
grant PHY-0457014. Helpful discussions and the support of Heiko Rohdjess
is gratefully acknowledged. The authors thank W.~J.~Briscoe and I.~I.~Strakovsky for 
making the files of Sp07 available for use in this manuscript. One of the authors (Ch.E.)
acknowledges inspiring discussions with T.-S.H. Lee and the hospitality of the ANL Theory
Group during the final stages of the manuscript.


\bibliographystyle{apsrev}

\clearpage

\begin{table}
\begin{tabular}{|c|c|} \hline \hline
{\rm Threshold} & $T_{\rm lab}$ [MeV] \\
\hline \hline
$NN\pi$         &  285 \\
$N\Delta$       &  632 \\
$NN^*(1440)$    & 1136 \\
$NN\eta$        & 1256 \\
$\Delta\Delta$  & 1355 \\
$NN^*(1535)$    & 1381 \\
\hline \hline
\end{tabular}
\caption{Laboratory (kinetic) energy thresholds for single-pion and single-eta 
production in the $NN$ system. In case of the resonances, the number indicates the
laboratory kinetic energy for which the resonance becomes on-shell.}
\label{table-1}
\end{table}

\begin{table}
\begin{tabular}{|c|c|c|ccc|} \hline\hline
 Resonance & Channel & $\Gamma_{{\rm{partial}}}$ [MeV] & R [GeV$^{-1}$] & 
$\gamma$ & $\kappa$ [MeV] \\
\hline\hline
  $\Delta$ & $N\pi$ & $\sim$ 120$^{(a)}$ & 6.3  & 0.749 & 200 \\
 $N^*(1440)$ & $N\pi$ & $\sim$ 270$^{(b)}$ &6.3 & 0.75  & 400 \\
$N^*(1535)$  & $N\pi$ & $\sim$ 60$^{(a)}$ &6.3 & 0.1286 & 400 \\
$N^*(1535)$  & $N\eta$& $\sim$ 79$^{(a)}$ &6.3 & 0.4284  & 400 \\
\hline\hline
\end{tabular}
\caption{The parameters for the description of the partial resonance 
width as given in Eq.~(\ref{eq:3.8}). 
The superscript $(a)$ indicates values taken from Ref.~\cite{PDG}, $(b)$ 
indicates the value from Ref.~\cite{Manley}. 
}
\label{table-2}
\end{table}

\vspace*{10mm}

\begin{table}
\begin{tabular}{|c|c|cc|c|cc|c|}  \hline \hline
Vertex & $\alpha$ &  $J^P$ & I & $m_\alpha$ & $\frac{g^2_{NN\alpha}}{4\pi}$&
$f_\alpha/g_\alpha$ & $\Lambda_\alpha$ [MeV] \\
\hline
$NN\alpha$ & $\pi$        & $0^-$& 1 & 138.03 & 13.8 &    &  1519 \\
           & $\eta$       & $0^-$& 0 & 547.3  &  5.81 &    & 830  \\
           & $\rho$       & $1^-$& 1 & 769.   &  1.1 & 5.9& 1281  \\
           & $\omega$     & $1^-$& 0 & 782.6  & 23.1&  0. & 1382  \\
           & $\sigma$~($T=1$)& $0^+$& 0 &497.   & 5.505 &    & 1807  \\
           & $\sigma$~($T=0$)& $0^+$& 0 &480.   & 3.545&    & 1990  \\
           & $a_0$         & $0^+$& 1 &980.   & 4.75 &    & 1004  \\
           & $\eta'$       & $0^-$& 0 &958.   & 1.62 &    & 1433  \\
           & $f_0  $       & $0^+$& 0 &980.   & 2.5  &    & 1274  \\
\hline
           &     & &          &      & $\frac{f^2_{N\Delta\alpha}}{4\pi}$ &
 & $\Lambda_{N\Delta\alpha}$ [MeV] \\
\hline
$N\Delta\alpha$ & $\pi$  & $0^-$& 1 & 138.03 & 0.224 &  & 640 \\
                & $\rho$ & $1^-$& 1 & 769.   & 20.45 &  & 1508 \\
\hline\hline
\end{tabular}
\caption{Meson parameters of the model (A) including the $\Delta$ isobar 
as the only mechanism for pion production. 
$J$, $P$, and $I$ denote spin, parity and isospin of the meson. 
Each meson vertex is multiplied by a form factor as given in
Eq.~(\ref{eq:dipolecut}), where $n_\alpha$=~1 except for the $NN\rho$ vertex 
where $n_\alpha$=~2.  
}
\label{table-3}
\end{table}

\vspace*{20mm} 

\begin{table}
\begin{tabular}{|c|c|cc|c|cc|c|}  \hline \hline
Vertex & $\alpha$ &  $J^P$ & I & $m_\alpha$ & $\frac{g^2_{NN\alpha}}{4\pi}$&
$f_\alpha/g_\alpha$ & $\Lambda_\alpha$ [MeV] \\
\hline
$NN\alpha$ & $\pi$        & $0^-$& 1 & 138.03 & 13.8 &    &  1498 \\
           & $\eta$       & $0^-$& 0 & 547.3  &  2.0 &    & 1300  \\
           & $\rho$       & $1^-$& 1 & 769.   &  1.1 & 5.9& 1281  \\
           & $\omega$     & $1^-$& 0 & 782.6  & 23.6&  0. & 1396  \\
           & $\sigma$~($T=1$)& $0^+$& 0 &494.   & 5.195 &    & 1924  \\
           & $\sigma$~($T=0$)& $0^+$& 0 &495.   & 3.603&    & 1700  \\
           & $a_0$         & $0^+$& 1 &980.   & 4.75 &    & 1004  \\
           & $\eta'$       & $0^-$& 0 &958.   & 1.24 &    & 1433  \\
           & $f_0  $       & $0^+$& 0 &980.   & 2.5  &    & 1471  \\
\hline
           &     & &          &      & $\frac{f^2_{N\Delta\alpha}}{4\pi}$ &
 & $\Lambda_{N\Delta\alpha}$ [MeV] \\
\hline
$N\Delta\alpha$ & $\pi$  & $0^-$& 1 & 138.03 & 0.224 &  & 677 \\
                & $\rho$ & $1^-$& 1 & 769.   & 20.45 &  & 1508 \\
\hline
                &   & &    &  & $\frac{f^2_{NN^*\pi}}{4\pi}$ 
 & $\frac{g^2_{NN^*\sigma}}{4\pi}$
 & $\Lambda_{NN^*\alpha}$ [MeV] \\
\hline
$NN^*(1440)\alpha$ & $\pi$  & $0^-$& 1 & 138.03 & 0.023 &  & 800 \\
                & $\sigma$ & $0^+$& 1 & 494.   &  & 0.74  & 1119 \\
\hline\hline
\end{tabular}
\caption{Meson parameters of the model (B) including  $\Delta$ isobar and the
$N^*(1440)$
as mechanism for pion production. $J$, $P$, and $I$ denote spin, parity and
isospin of the meson.
Each meson vertex is multiplied with a form factor as given in
Eq.~(\ref{eq:dipolecut}), where $n_\alpha$=~1 except for the $NN\rho$ vertex
where $n_\alpha$=~2.
}
\label{table-4}
\end{table}

\vspace*{20mm}

\begin{table}
\begin{tabular}{|c|c|cc|c|cc|c|}  \hline \hline
Vertex & $\alpha$ &  $J^P$ & I & $m_\alpha$ & $\frac{g^2_{NN\alpha}}{4\pi}$&
$f_\alpha/g_\alpha$ & $\Lambda_\alpha$ [MeV] \\
\hline
$NN\alpha$ & $\pi$        & $0^-$& 1 & 138.03 & 13.8 &    &  1429 \\
           & $\eta$       & $0^-$& 0 & 547.3  &  2.0 &    & 1298  \\
           & $\rho$       & $1^-$& 1 & 769.   &  0.97 & 5.9& 1310  \\
           & $\omega$     & $1^-$& 0 & 782.6  & 20.5&  0. & 1421  \\
           & $\sigma$~($T=1$)& $0^+$& 0 &503.   & 5.254 &    & 2000  \\
           & $\sigma$~($T=0$)& $0^+$& 0 &530.   & 4.164&    & 1516  \\
           & $a_0$         & $0^+$& 1 &980.   & 4.75 &    & 968  \\
           & $\eta'$       & $0^-$& 0 &958.   & 1.57 &    & 1424  \\
           & $f_0  $       & $0^+$& 0 &980.   & 2.31  &    & 1442  \\
\hline
           &     & &          &      & $\frac{f^2_{N\Delta\alpha}}{4\pi}$ &
 & $\Lambda_{N\Delta\alpha}$ [MeV] \\
\hline
$N\Delta\alpha$ & $\pi$  & $0^-$& 1 & 138.03 & 0.224 &  & 618 \\
                & $\rho$ & $1^-$& 1 & 769.   & 20.45 &  & 1422 \\
\hline
       &   & &    &  & $\frac{f^2_{NN^*\pi}}{4\pi}$ &
$\frac{g^2_{NN^*\sigma}}{4\pi}$
 & $\Lambda_{NN^*\alpha}$ [MeV] \\
\hline
$NN^*(1440)\alpha$ & $\pi$  & $0^-$& 1 & 138.03 & 0.023 &  & 800 \\
                & $\sigma$ & $0^+$& 0 & 503.   &  & 0.74  & 1119 \\
\hline
       &   & &    &  & $\frac{g^2_{NN^*\alpha}}{4\pi}$ &
 & $\Lambda_{NN^*\alpha}$ [MeV] \\
\hline
$NN^*(1535)\alpha$ & $\pi$  & $0^-$& 1 & 138.03 & 0.05 &  & 809 \\
                & $\eta$ & $0^-$& 0 & 547.3.   &  0.99 &  & 962 \\
\hline\hline
\end{tabular}
\caption{Meson parameters of the model (C) including  $\Delta$ isobar and the
nucleon resonances $N^*(1440)$ and $N^*(1535)$
as mechanism for pion production. $J$, $P$, and $I$ denote spin, parity and
isospin of the meson.
Each meson vertex is multiplied with a form factor as given in
Eq.~(\ref{eq:dipolecut}), where $n_\alpha$=~1 except for the $NN\rho$ vertex
where $n_\alpha$=~2.
}
\label{table-5}
\end{table}

\clearpage


\noindent

\vspace{3cm}

\begin{figure}
\end{figure}

\noindent

\begin{figure}
\begin{center}
\includegraphics[width=10cm]{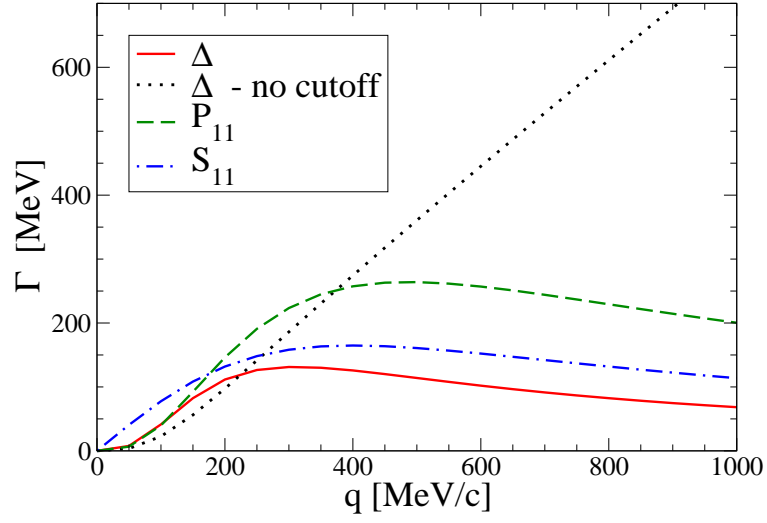}
\end{center}
\caption{The width of the resonances $\Delta(1232)$ (solid line), $N^*(1440)$ 
(dashed line) , and $N^*(1535)$ (dash-dotted line) as function of the c.m.\
momentum of the resonance. 
The dotted line indicates the width of the $\Delta$ resonance when no 
cutoff $\kappa$ is applied. 
}
\label{fig2}
\end{figure}

\noindent

\begin{figure}
\begin{center}
\includegraphics[width=12cm]{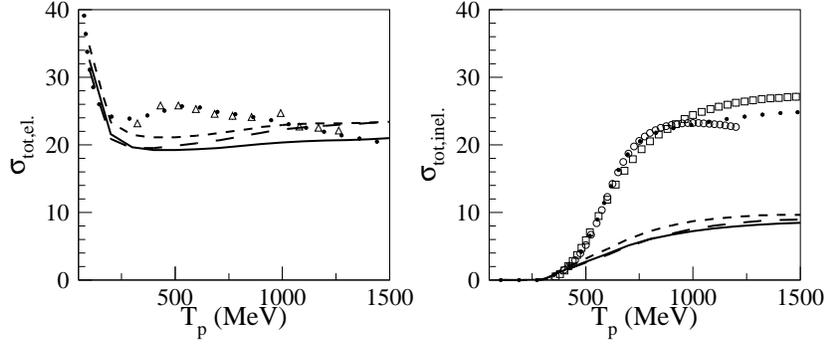}
\end{center}
\caption{The total cross section for  elastic (left) and inelastic (right) $pp$
scattering as function of the laboratory projectile kinetic energy.
The solid line gives the prediction of the model C (Table~\ref{table-5}),
containing the contributions of all three resonances $P_{33}$, $P_{11}$, and 
$S_{11}$.
The short-dashed line is based on the model B (Table~\ref{table-4})
containing only the resonances $P_{33}$ and  $P_{11}$.  The 
long-dashed line represents the results of model A (Table~\ref{table-3})
containing only the $P_{33}$ resonance in addition to the one-meson
contributions. 
The SAID analysis Sp07 \protect\cite{Arndt:2007qn} is indicated by the 
dotted line.
The experimental data are referenced in the SAID database~\protect\cite{SAID}. 
}
\label{fig3}
\end{figure}

\noindent

\begin{figure}
\begin{center}
\includegraphics[width=12cm]{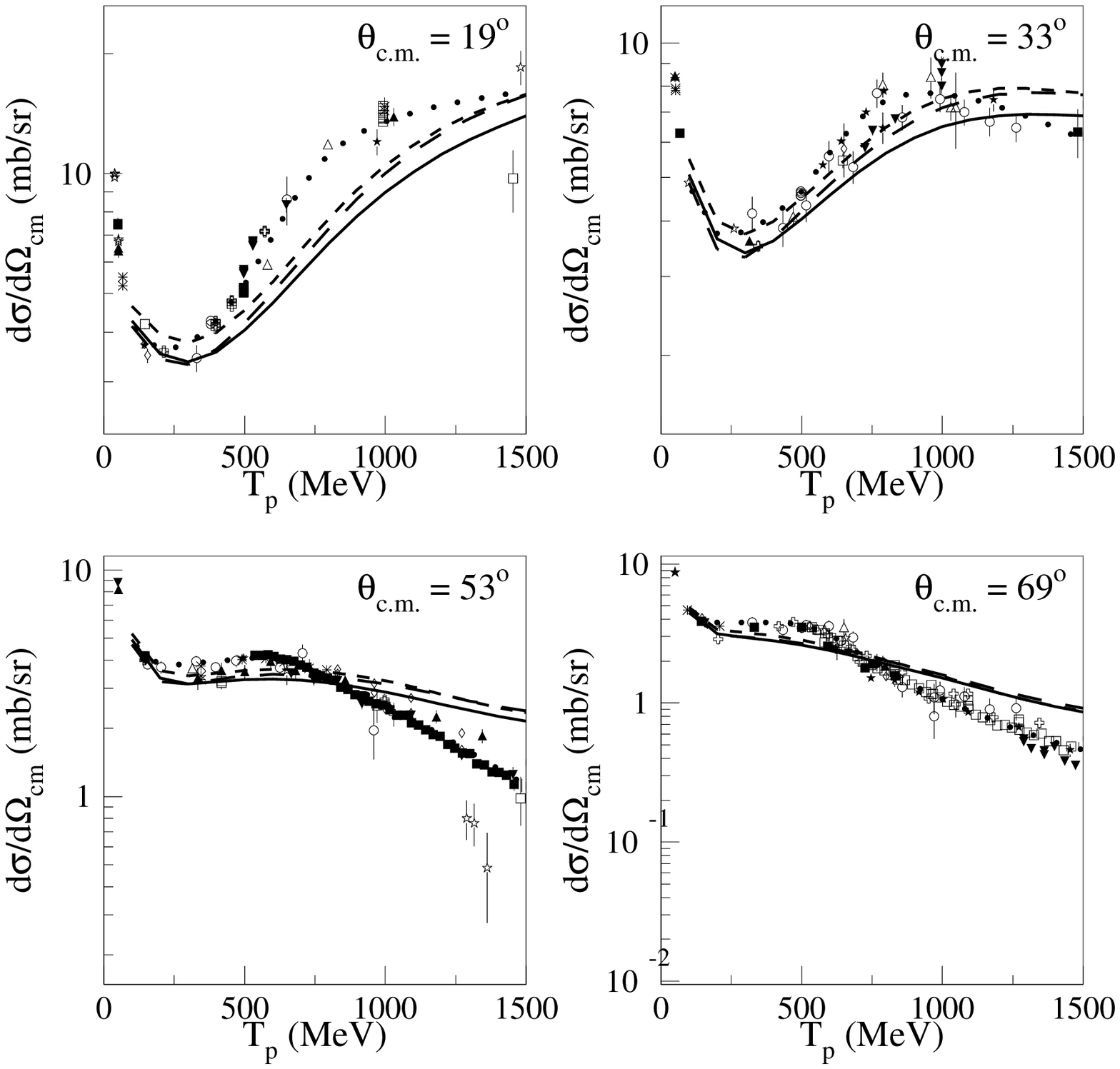}
\end{center}
\caption{The differential c.m.\ cross section for $pp$ scattering as function of 
the laboratory projectile kinetic energy at selected angles. 
The notation of the curves is the same as in Fig.~\protect\ref{fig3}. 
The experimental data are referenced in the SAID database~\protect\cite{SAID}.
}
\label{fig4}
\end{figure}

\noindent

\begin{figure}
\begin{center}
\includegraphics[width=12cm]{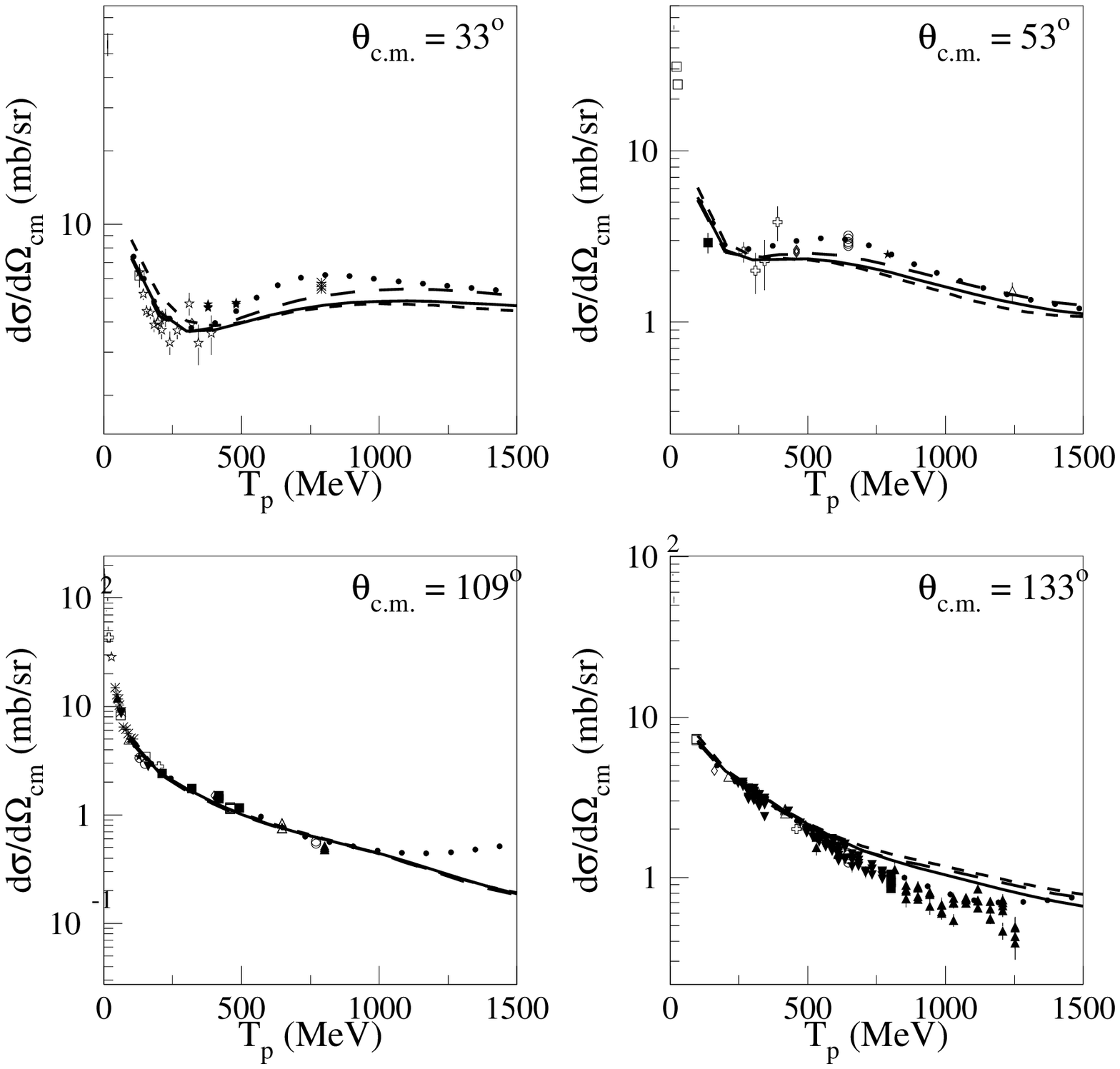}
\end{center}
\caption{The differential cross section for $np$ scattering as function of
the laboratory projectile kinetic energy at selected angles. 
The notation of the curves is the same as in Fig.~\protect\ref{fig3}. 
The experimental data are referenced in the SAID database~\protect\cite{SAID}.
}
\label{fig5}
\end{figure}

\noindent

\begin{figure}
\begin{center}
\includegraphics[width=12cm]{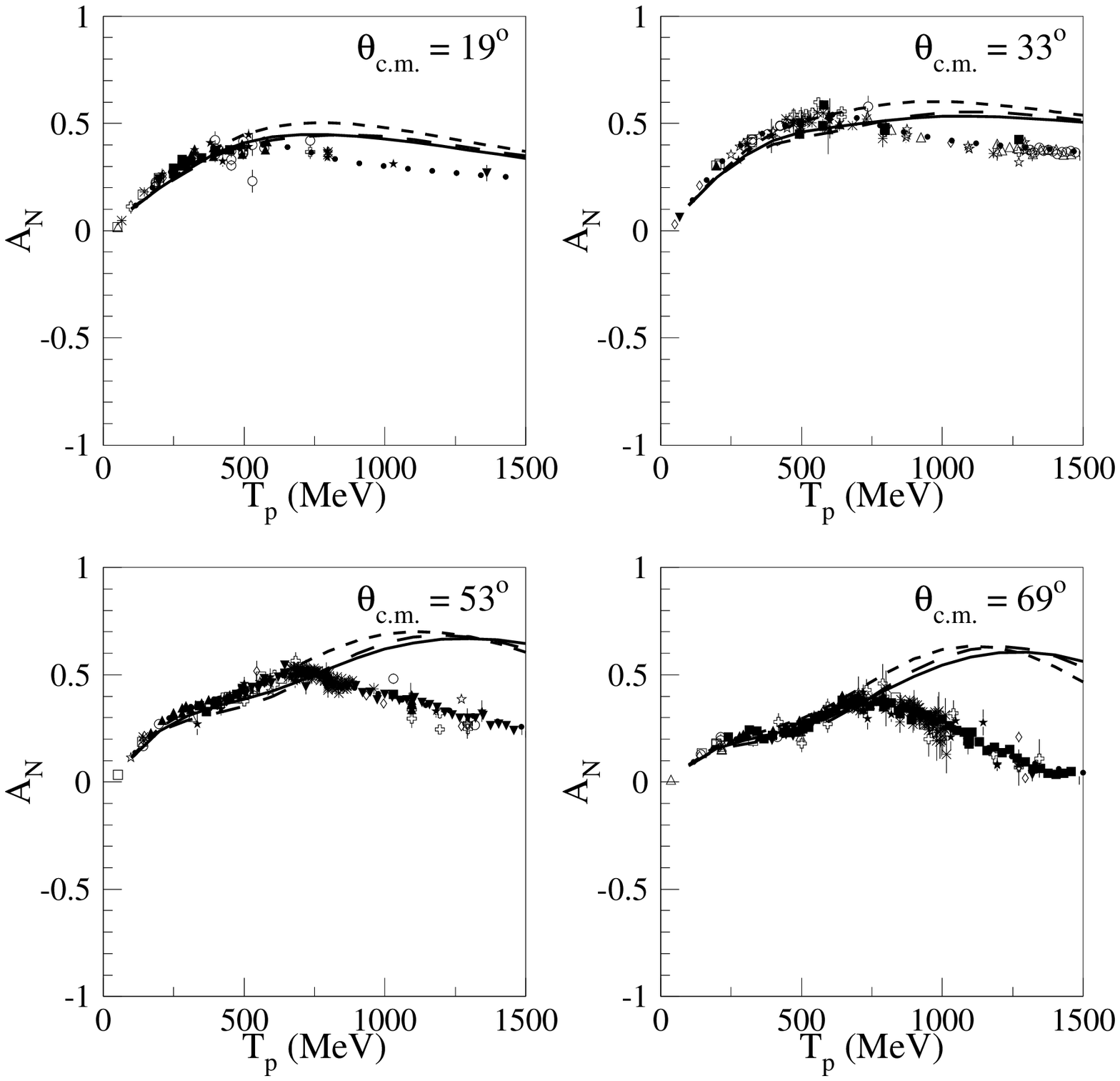}
\end{center}
\caption{The analyzing power $A_N$ for $pp$ scattering as function of
the laboratory projectile kinetic energy at selected angles. 
The notation of the curves is the same as in Fig.~\protect\ref{fig3}. 
The experimental data are referenced in the SAID database~\protect\cite{SAID}.
}
\label{fig6}
\end{figure}

\noindent
\begin{figure}
\begin{center}
\includegraphics[width=12cm]{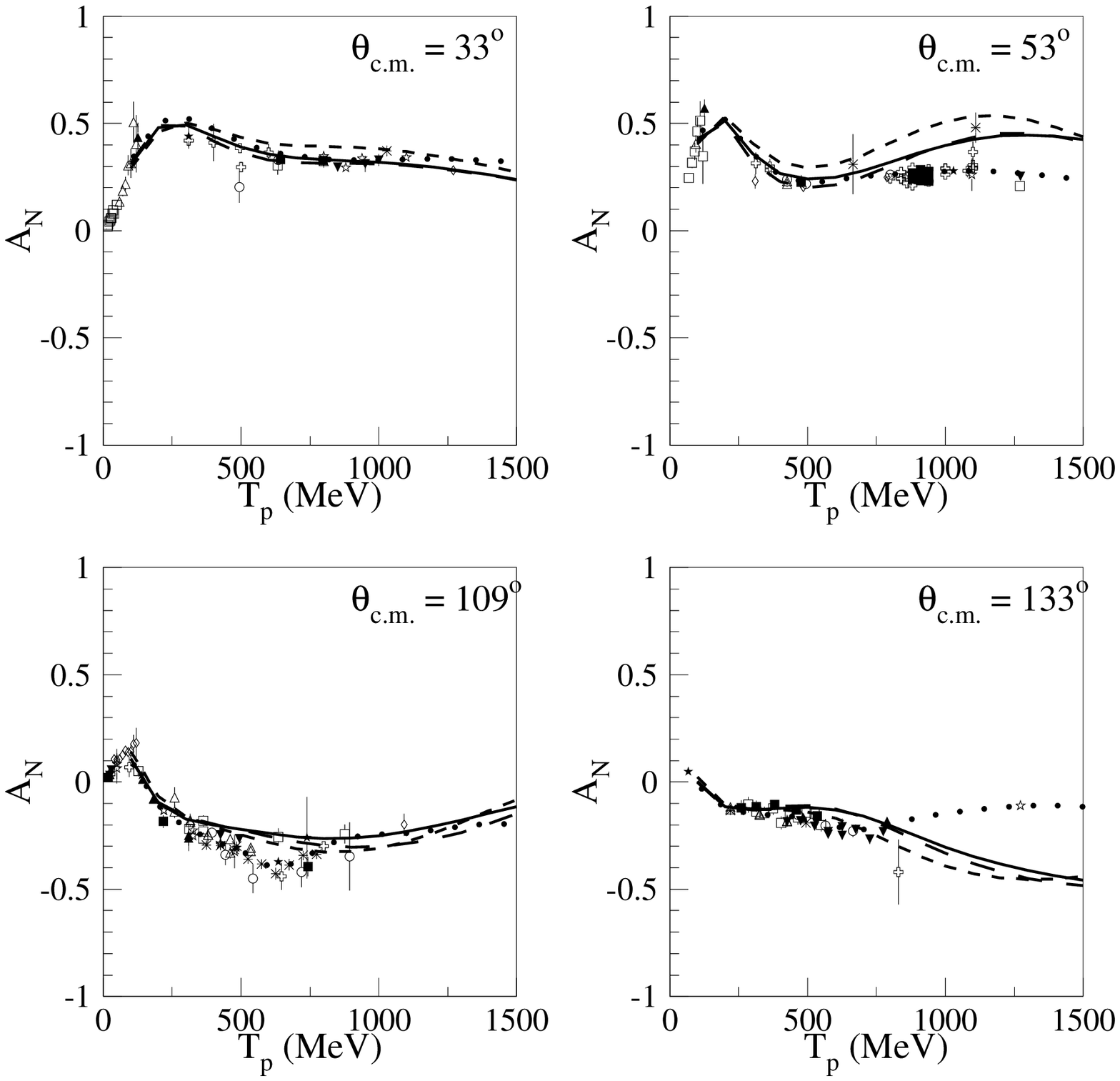}
\end{center}
\caption{The analyzing power $A_N$ for $np$ scattering as function of
the laboratory projectile kinetic energy at selected angles. 
The notation of the curves is the same as in Fig.~\protect\ref{fig3}. 
The experimental data are referenced in the SAID database~\protect\cite{SAID}.
}
\label{fig7}
\end{figure}

\noindent
\begin{figure}
\begin{center}
\includegraphics[width=12cm]{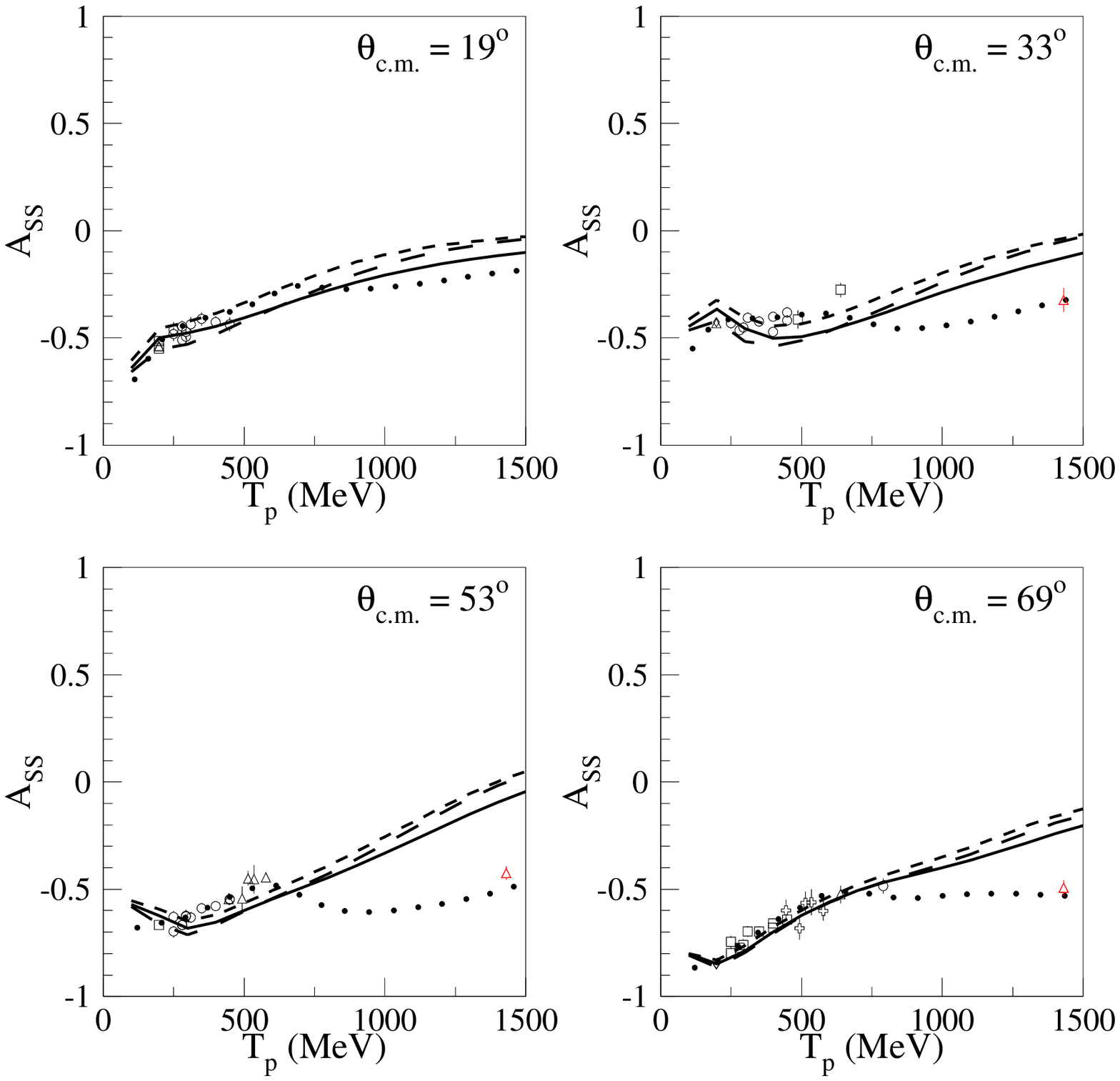}
\end{center}
\caption{The spin-correlation parameter $A_{SS}$ for $pp$ scattering as 
function of the laboratory projectile kinetic energy at selected angles. 
The notation of the curves is the same as in Fig.~\protect\ref{fig3}. 
The experimental data are referenced in the SAID database~\protect\cite{SAID}.
}
\label{fig8}
\end{figure}

\noindent
\begin{figure}
\begin{center}
\includegraphics[width=12cm]{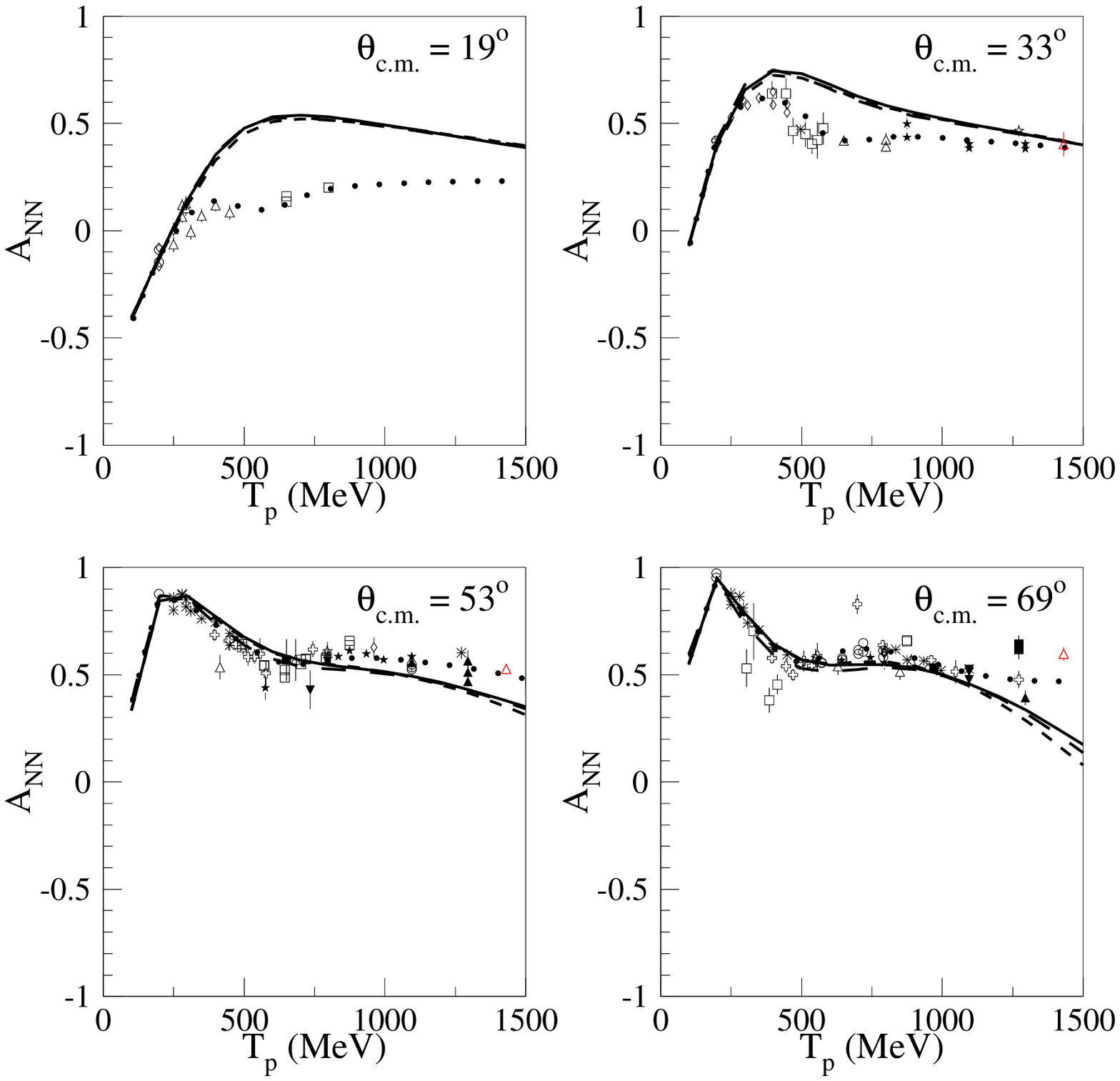}
\end{center}
\caption{The spin correlation parameter $A_{NN}$ for $pp$ scattering as 
function of the laboratory projectile kinetic energy at selected angles. 
The notation of the curves is the same as in Fig.~\protect\ref{fig3}. 
The experimental data are referenced in the SAID database~\protect\cite{SAID}.
}
\label{fig9}
\end{figure}

\noindent
\begin{figure}
\begin{center}
\includegraphics[width=12cm]{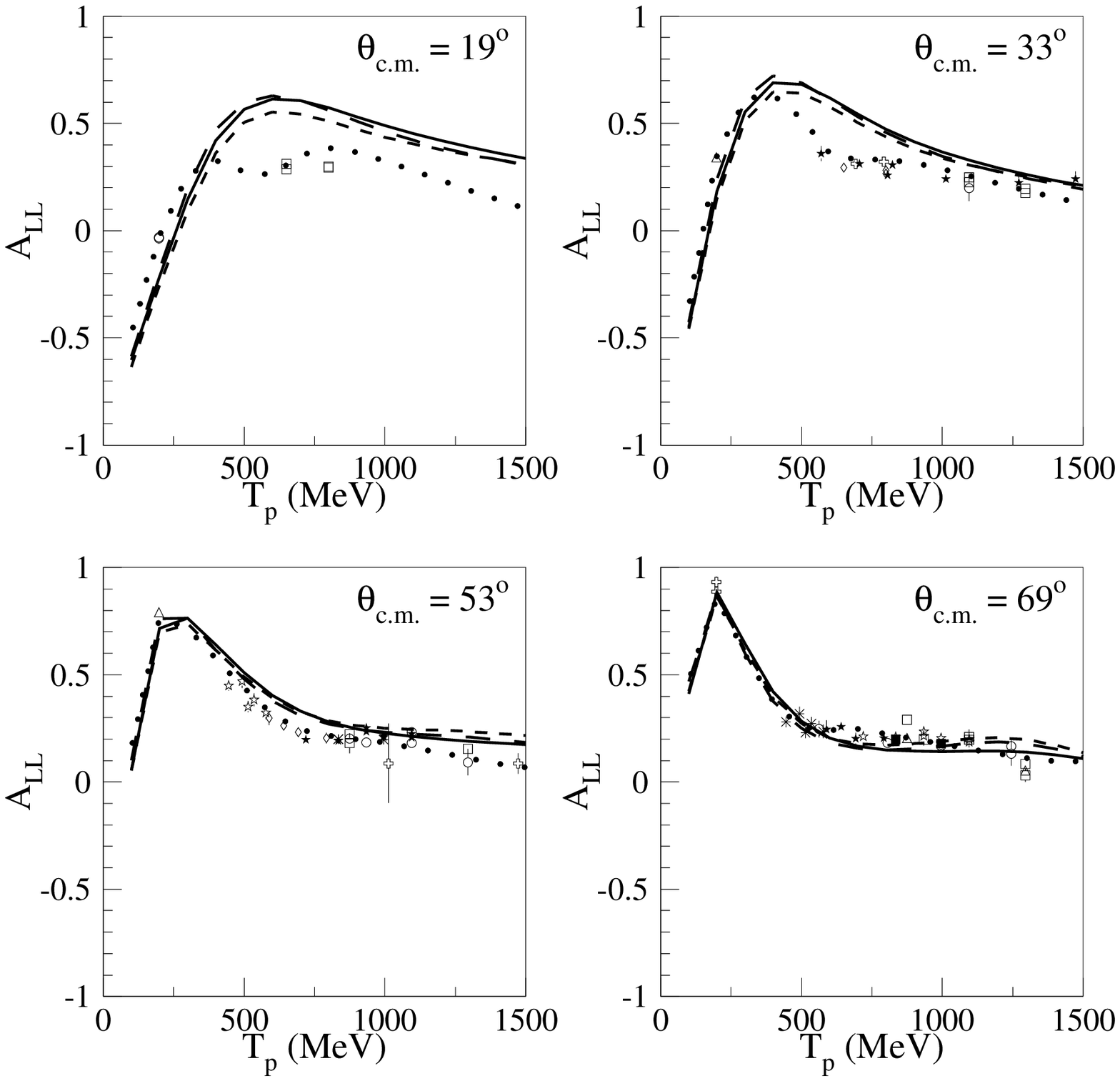}
\end{center}
\caption{The spin correlation parameter $A_{LL}$ for $pp$ scattering as 
function of the laboratory projectile kinetic energy at selected angles. 
The notation of the curves is the same as in Fig.~\protect\ref{fig3}. 
The experimental data are referenced in the SAID database~\protect\cite{SAID}.
}
\label{fig10}
\end{figure}

\noindent
\begin{figure}
\begin{center}
\includegraphics[width=12cm]{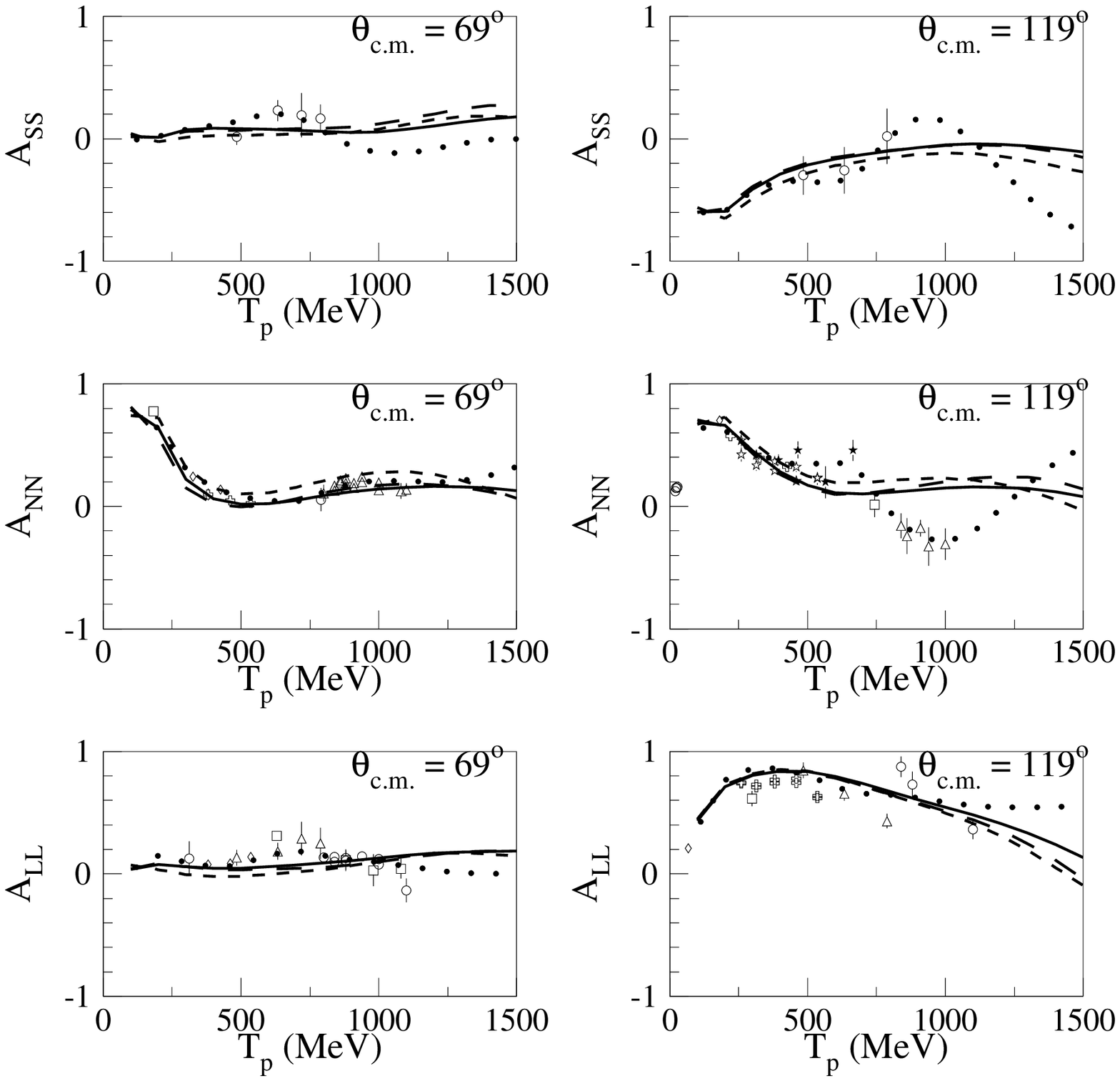}
\end{center}
\caption{The spin correlation parameters $A_{SS}$, $A_{NN}$, and $A_{LL}$  for 
$np$ scattering as function of the laboratory projectile kinetic energy at 
selected angles. 
The notation of the curves is the same as in Fig.~\protect\ref{fig3}. 
The experimental data are referenced in the SAID database~\protect\cite{SAID}.
}
\label{fig11}
\end{figure}

\begin{figure}
\begin{center}
\includegraphics[width=18cm]{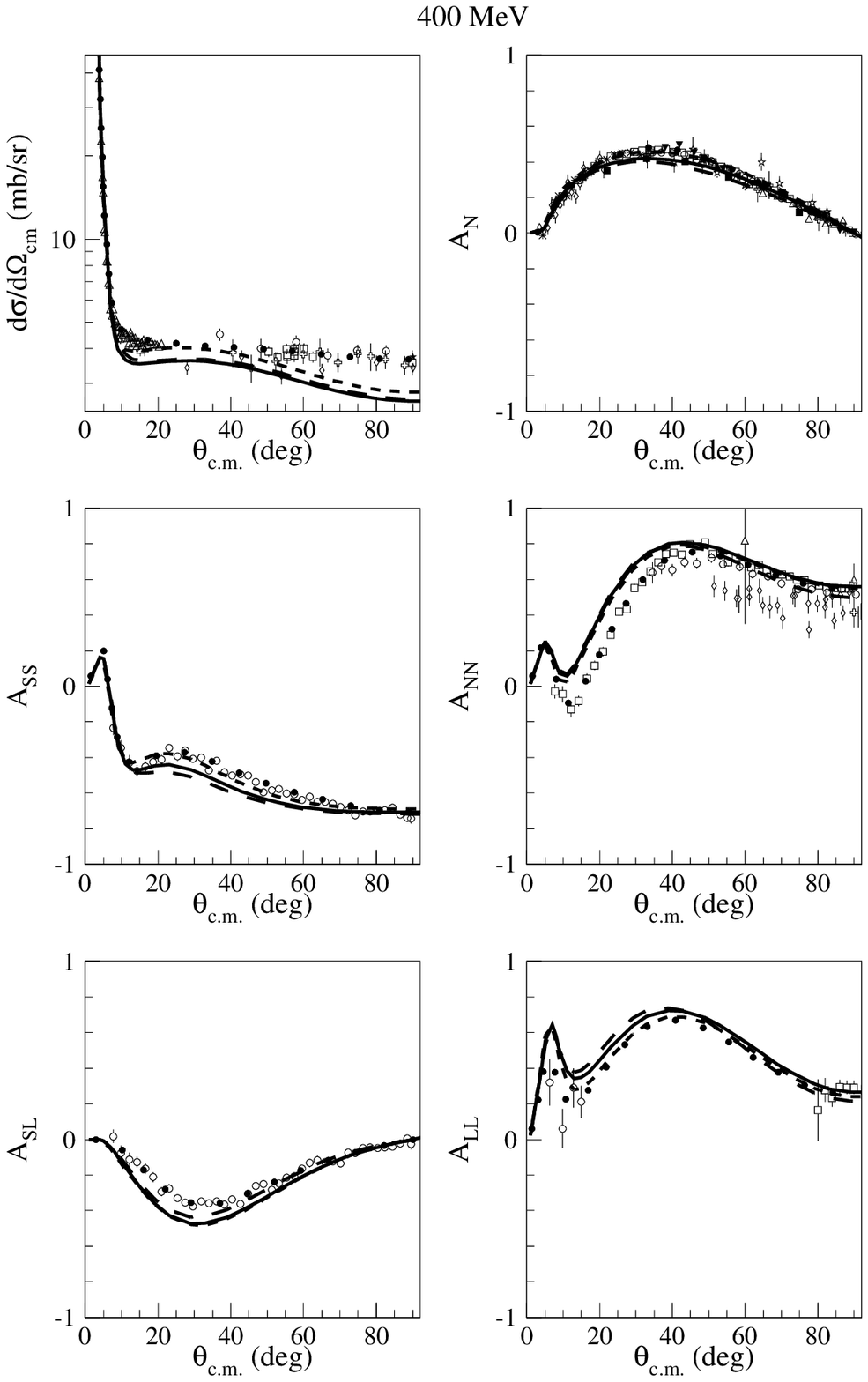}
\end{center}
\caption{The angular distribution of the differential cross section, the 
analyzing power and selected spin-correlation coefficients for $pp$ scattering 
at 400~MeV laboratory kinetic energy. 
The notation of the curves is the same as in Fig.~\protect\ref{fig3}. 
The experimental data are referenced in the SAID database~\protect\cite{SAID}.
}
\label{fig12}
\end{figure}

\begin{figure}
\begin{center}
\includegraphics[width=18cm]{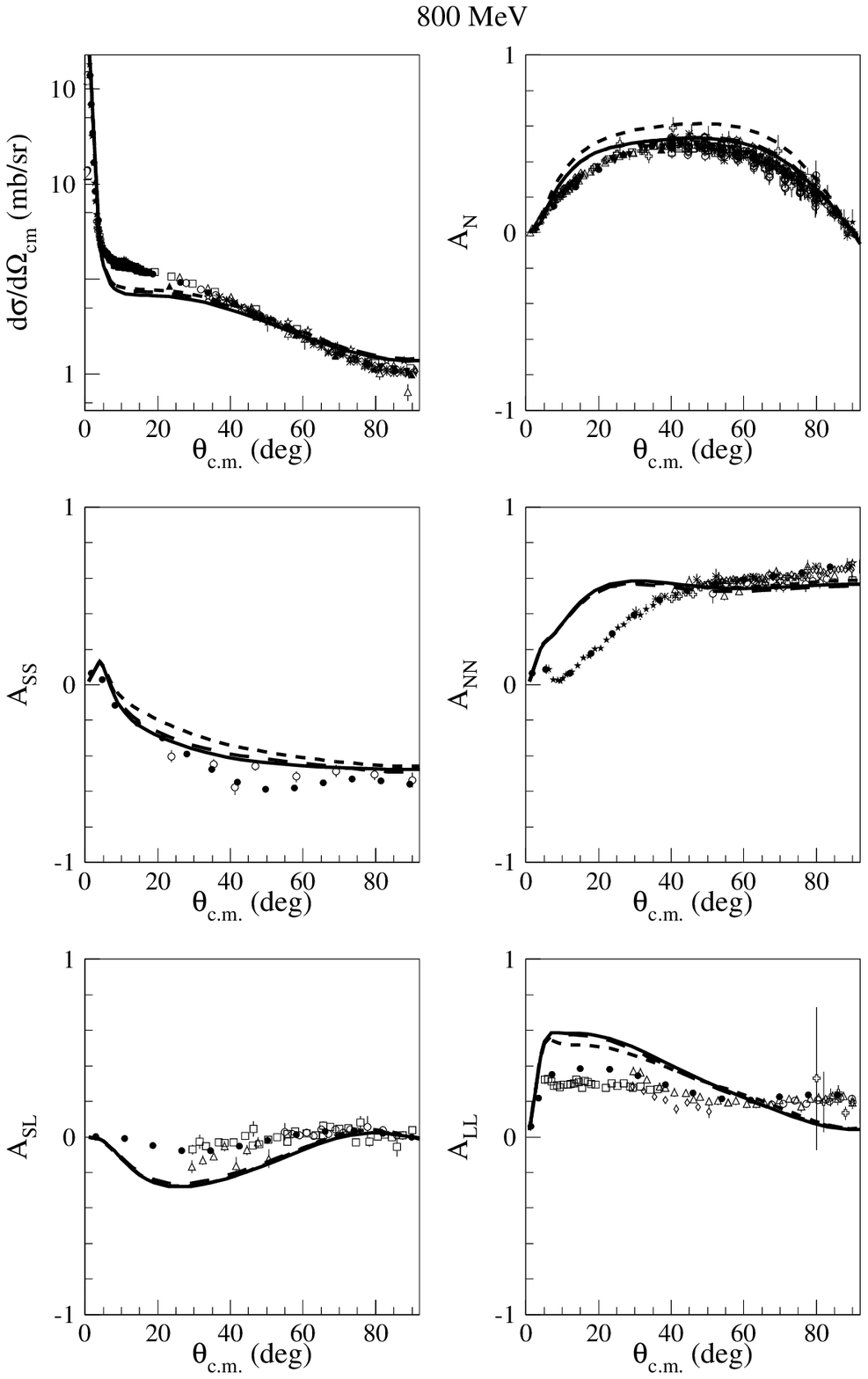}
\end{center}
\caption{The angular distribution of the differential cross section, the 
analyzing power and selected spin-correlation coefficients for $pp$ scattering 
at 800~MeV laboratory kinetic energy.
The notation of the curves is the same as in Fig.~\protect\ref{fig3}. 
The experimental data are referenced in the SAID database~\protect\cite{SAID}.
}
\label{fig13}
\end{figure}

\begin{figure}
\begin{center}
\includegraphics[width=18cm]{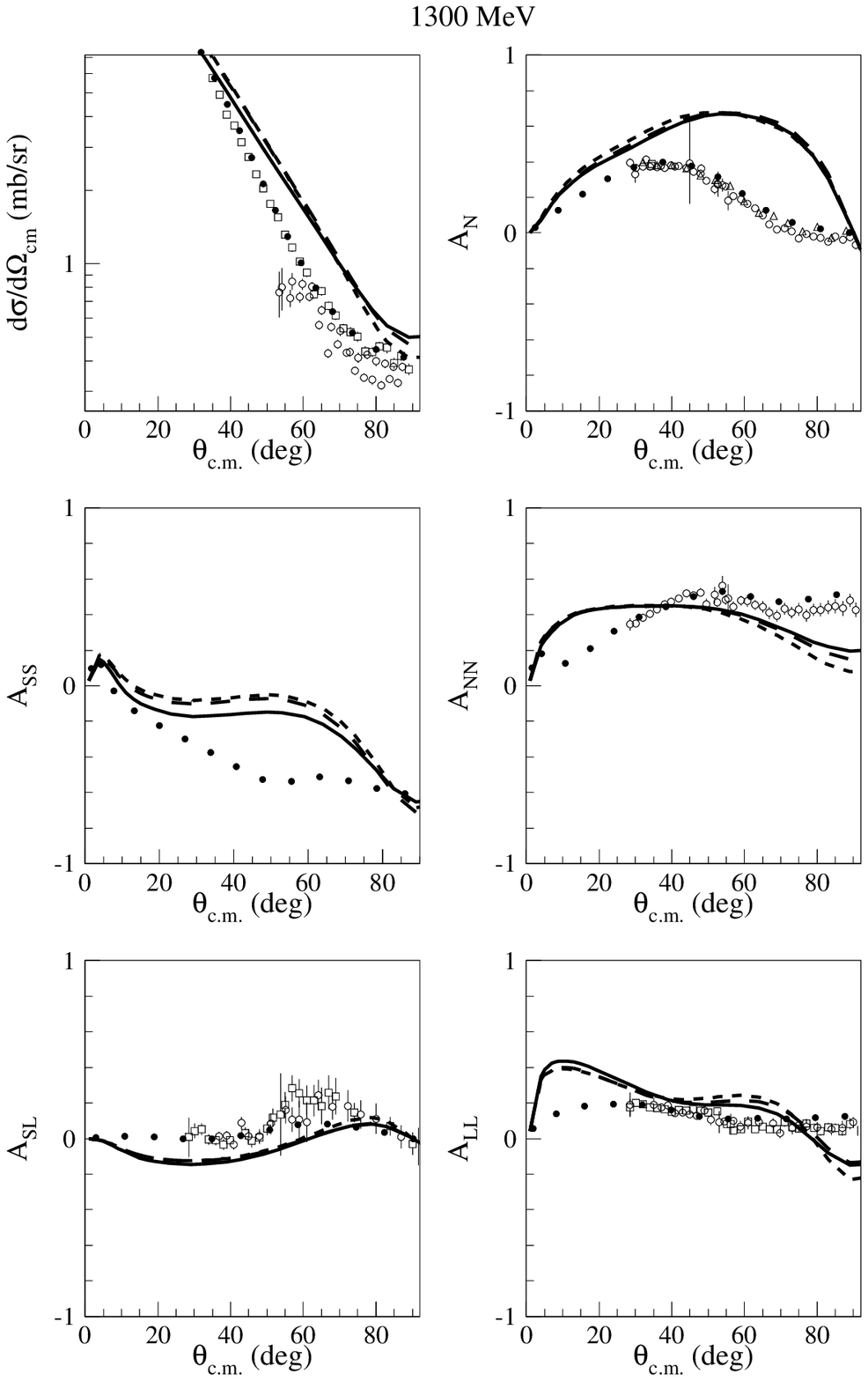}
\end{center}
\caption{The angular distribution of the differential cross section, the 
analyzing power and selected spin-correlation coefficients for $pp$ scattering 
at 1300~MeV laboratory kinetic energy.
The notation of the curves is the same as in Fig.~\protect\ref{fig3}. 
The experimental data are referenced in the SAID database~\protect\cite{SAID}.
}
\label{fig14}
\end{figure}

\begin{figure}
\begin{center}
\includegraphics[width=15cm,height=21cm]{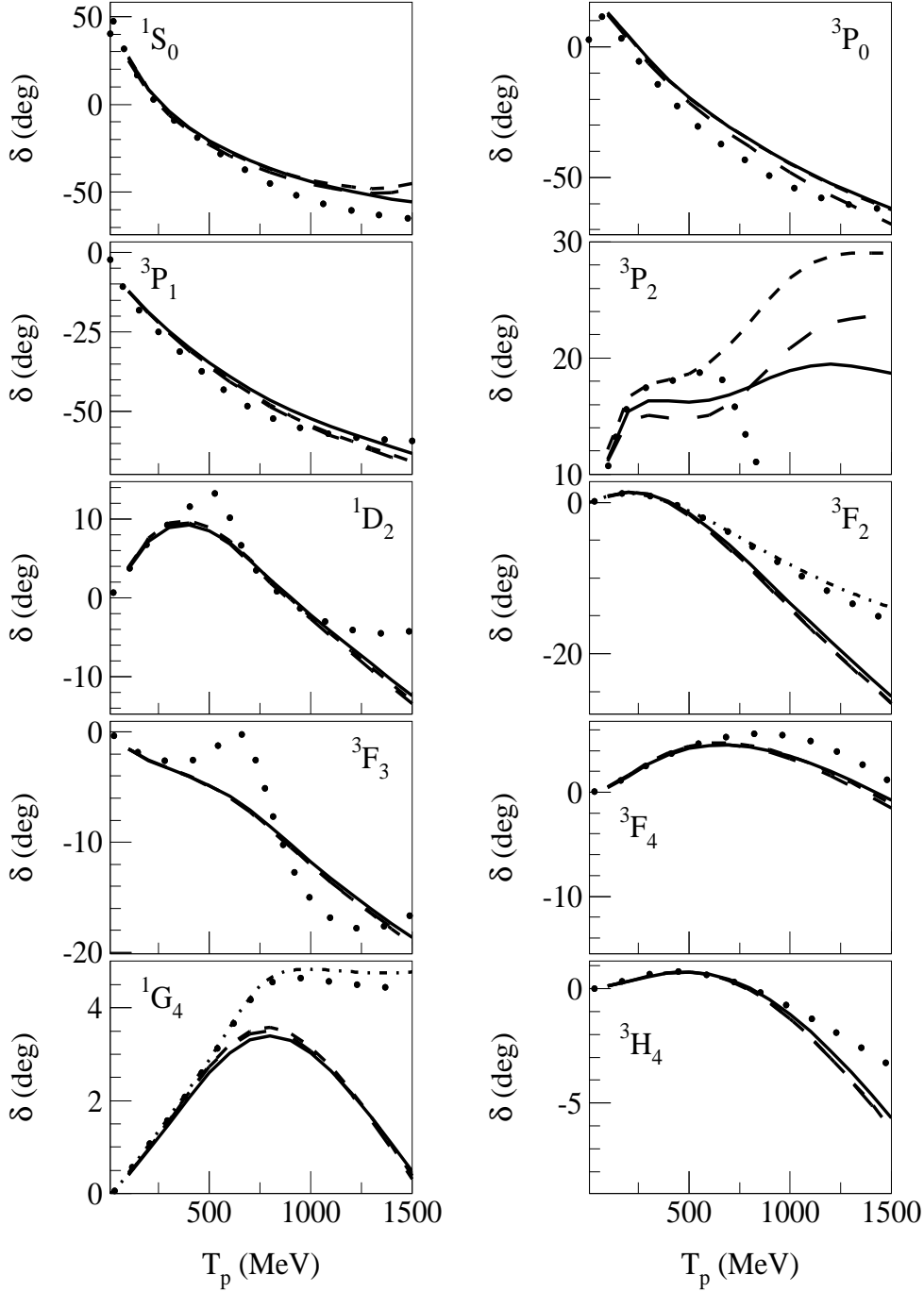}
\end{center}
\caption{The $T=1$ $NN$ partial wave 
phase shifts up to J=4 as function of the projectile 
laboratory kinetic energy. 
The solid line gives the prediction of the model C (Table~\ref{table-5}),
containing the contributions of the resonances $P_{33}$, $P_{11}$, and 
$S_{11}$.
The short-dashed line is based on the model B (Table~\ref{table-4})
containing only the resonances $P_{33}$ and  $P_{11}$.  The
long-dashed line represents the results of model A (Table~\ref{table-3})
containing only the $P_{33}$ resonance in addition to the one-meson
contributions.
The SAID analysis Sp07 \protect\cite{Arndt:2007qn} is indicated by the dotted 
line. The dash-dotted line in selected phase shifts (here $^3$F$_2$ and
$^1$G$_4$) represents the SAID analysis Sp04. }
\label{fig15}
\end{figure}

\noindent
\begin{figure}
\begin{center}
\includegraphics[width=15cm]{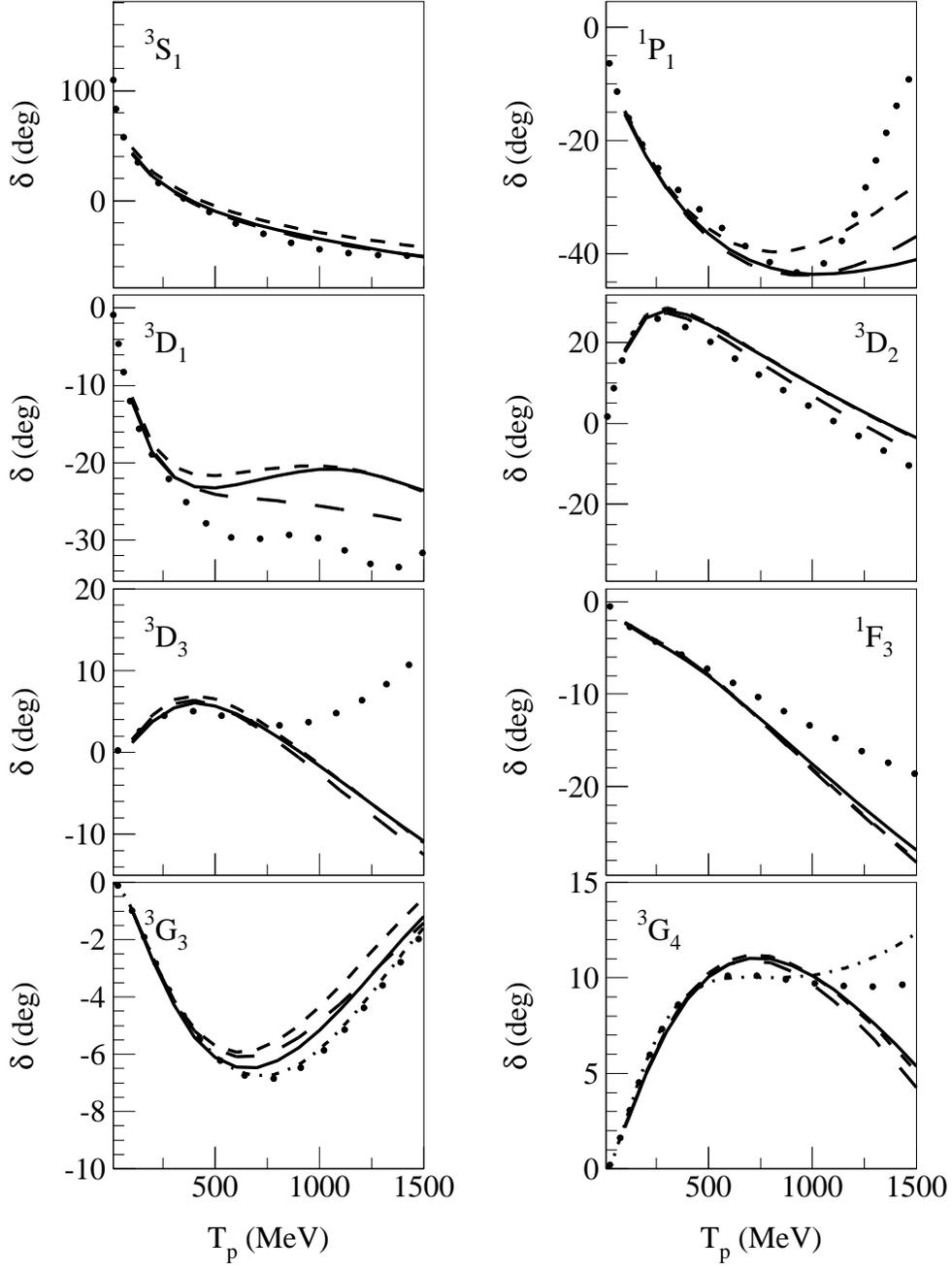}
\end{center}
\caption{The $T=0$ $NN$ partial wave
phase shifts up to J=4 as function of the projectile 
laboratory kinetic energy. 
The notation of the curves is the same as in Fig.~\protect\ref{fig15}.
}
\label{fig16}
\end{figure}

\begin{figure}
\begin{center}
\includegraphics[width=15cm]{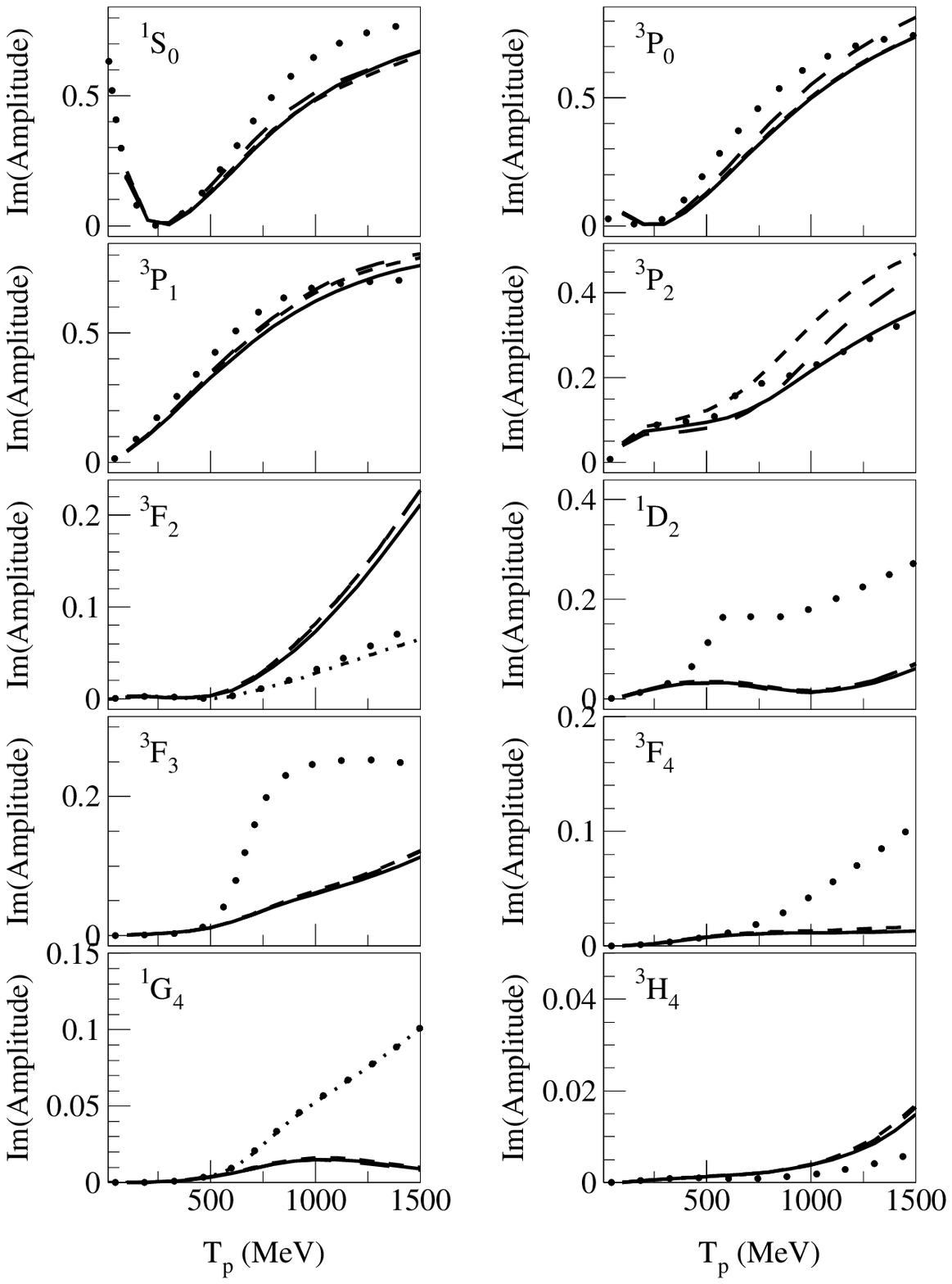}
\end{center}
\caption{The imaginary parts of the $T=1$ $NN$ partial wave amplitudes
up to J=4 as function of the projectile laboratory kinetic energy.
The notation of the curves is the same as in Fig.~\protect\ref{fig15}.
}
\label{fig17}
\end{figure}

\begin{figure}
\begin{center}
\includegraphics[width=15cm]{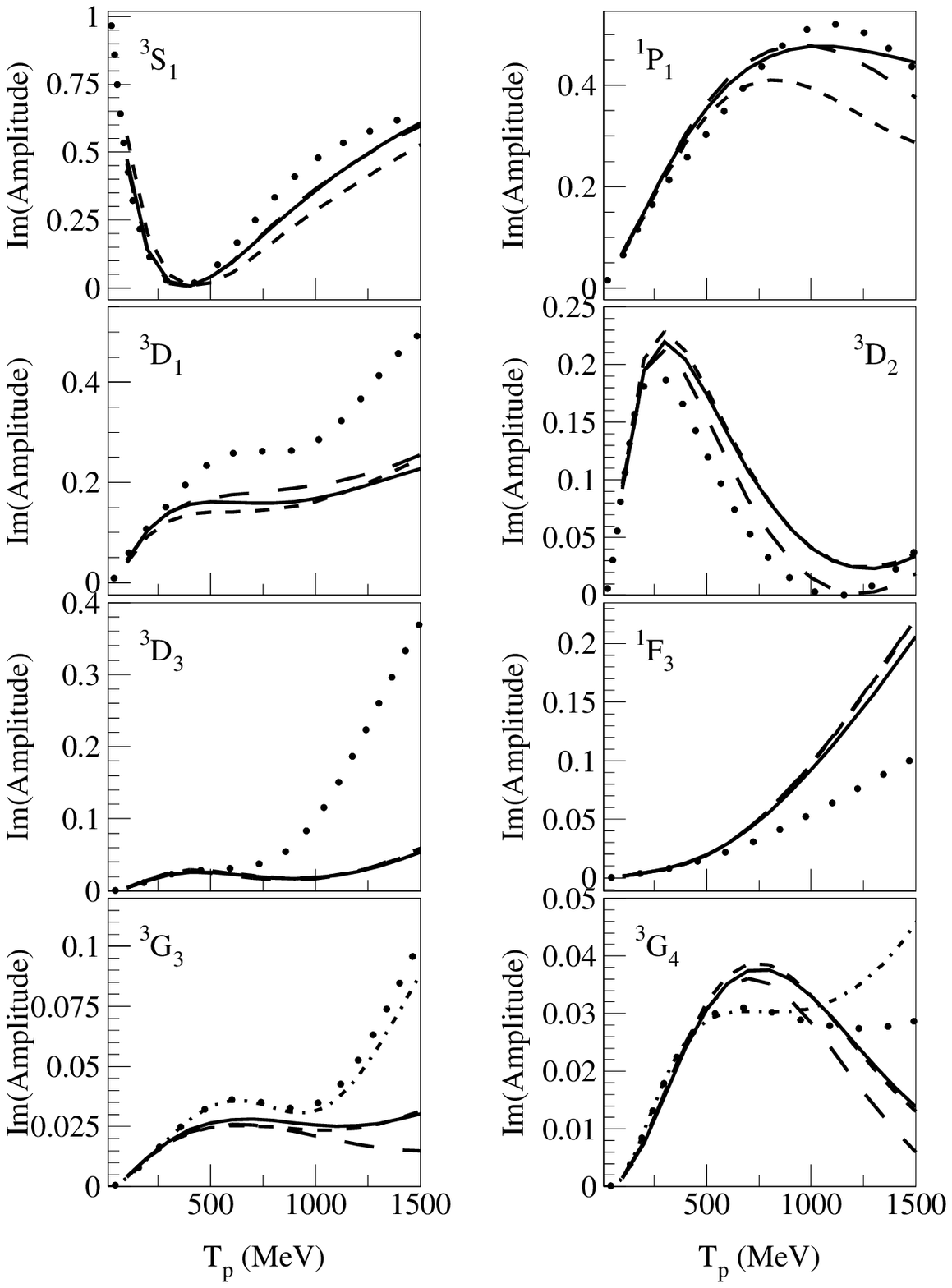}
\end{center}
\caption{The imaginary parts of the $T=0$ $NN$ partial wave amplitudes
 up to J=4 as function of the projectile laboratory kinetic energy.
The notation of the curves is the same as in Fig.~\protect\ref{fig15}.
}
\label{fig18}
\end{figure}

\begin{figure}
\begin{center}
\includegraphics[width=15cm]{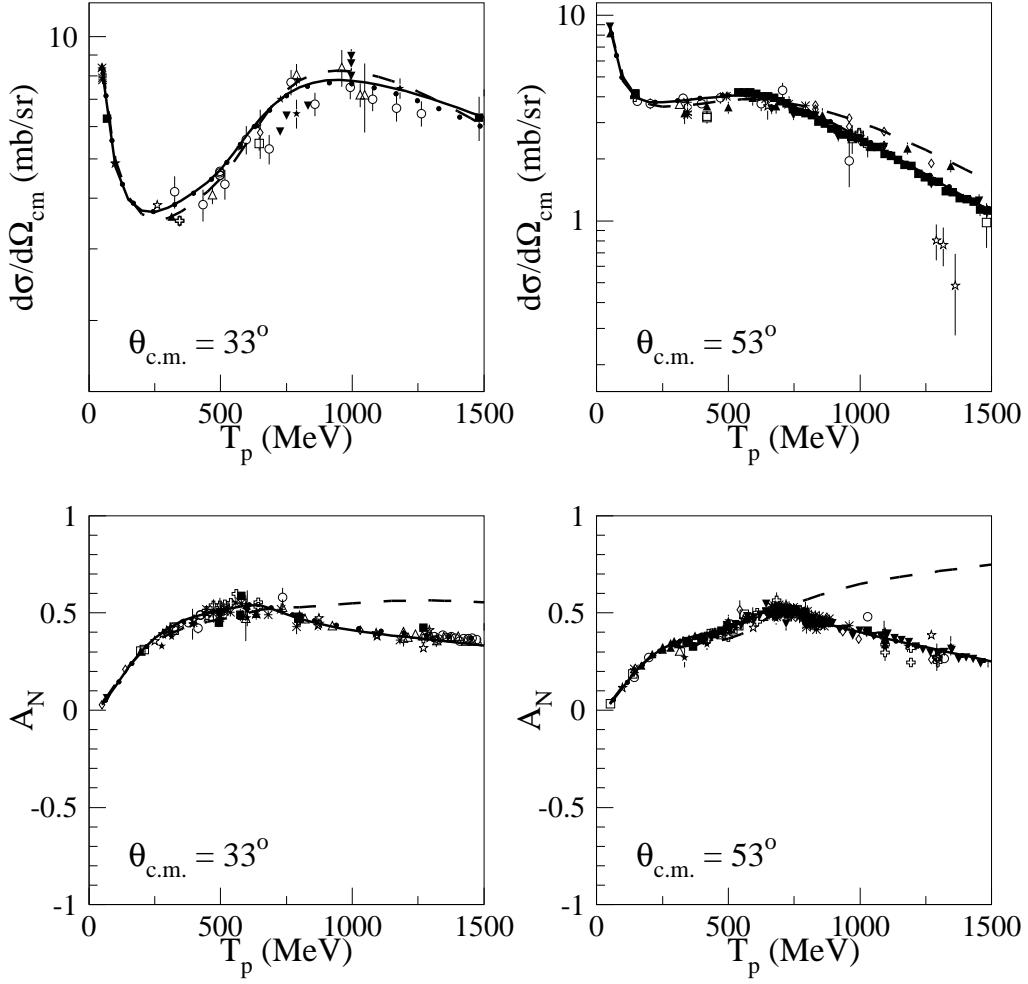}
\end{center}
\caption{The differential cross section (top panels) and the
analyzing power $A_N$ (bottom panels) for $pp$ scattering as function
of the laboratory projectile kinetic energy at two selected angles.
The dashed lines are based on the calculation with the parameters
of our model A (Table~\ref{table-3}). For the solid line the 
$^3$P$_2$ phase shift of the model is replaced by the SAID
analysis. The experimental data are referenced in the SAID
database~\protect\cite{SAID}.
}
\label{fig19}
\end{figure}

\begin{figure}
\begin{center}
\includegraphics[width=15cm]{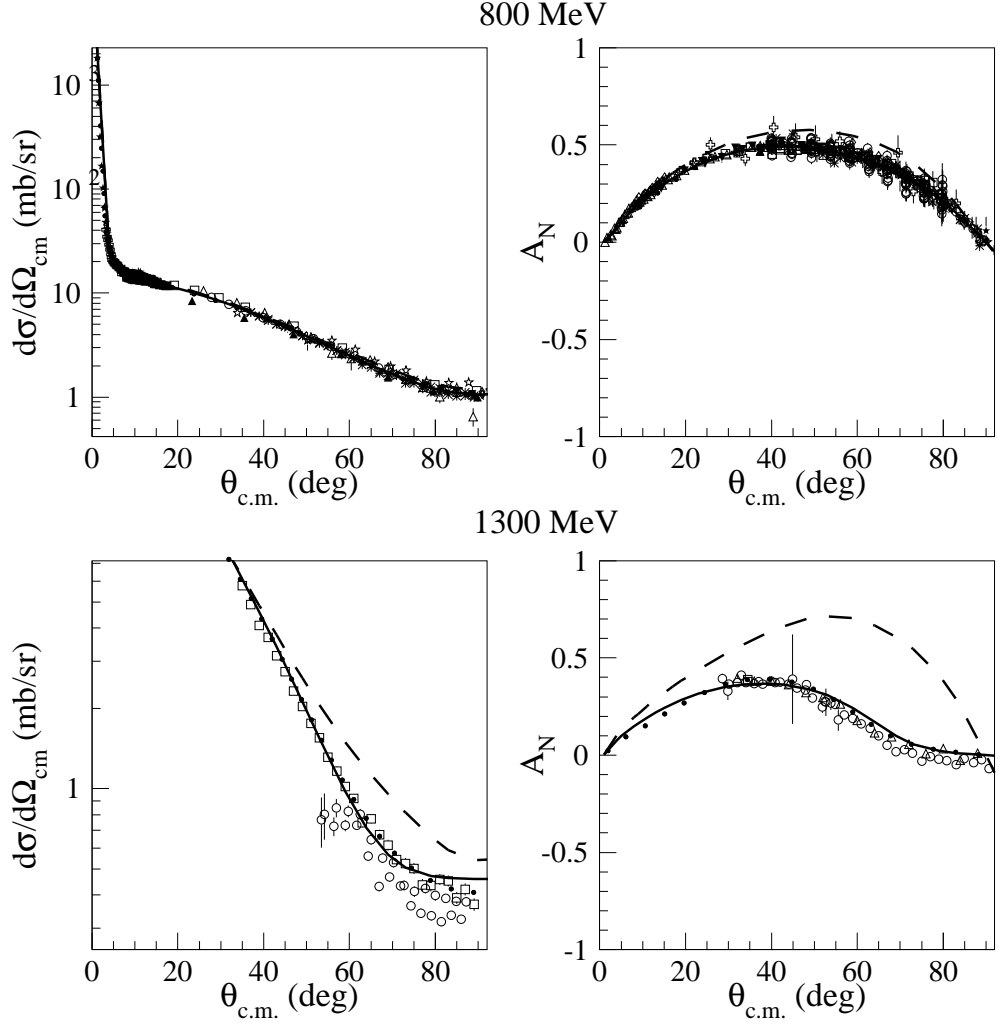}
\end{center}
\caption{The angular distribution of the differential cross section
(left panels) and the analyzing power for $pp$ scattering at 800~MeV
laboratory kinetic energy (top panels) and 1300~MeV (bottom panels).
The notation of the curves is the same as in
Fig.~\protect\ref{fig19}. The experimental
data are referenced in the SAID database~\protect\cite{SAID}.
}
\label{fig20}
\end{figure}

\end{document}